%% file: main.tex
\documentclass[manuscript]{acmart}

\AtBeginDocument{%
  \providecommand\BibTeX{{%
    \normalfont B\kern-0.5em{\scshape i\kern-0.25em b}\kern-0.8em\TeX}}}

\usepackage{algorithmicx}
\usepackage{tikz}
\usetikzlibrary{calc,angles,positioning,intersections,quotes,decorations.markings}
\usepackage{pgfplots}
\pgfplotsset{compat=1.16}
\tikzset{
	font={\fontsize{14pt}{12}\selectfont}}
\usepackage[noend]{algpseudocode}
\usepackage{amsmath,amsthm} 
\usepackage{array}
\usepackage{caption}
\usepackage{subcaption}
\usepackage{graphicx}
\usepackage{xcolor,colortbl}

\newcommand{\rt}[1]{\textcolor{red}{#1}}
\newcommand{\bt}[1]{{\color{cyan}{#1}}}
\newcommand{\ct}[1]{{\color{blue}{#1}}}
\newcommand{\mat}[1]{{\color{olive}{#1}}}

\usepackage{epstopdf}
\usepackage{setspace}
\usepackage{ctable}
\usepackage{romannum}
\usepackage{enumerate}
\usepackage{ragged2e}
\usepackage{multirow}
\usepackage{tabularx,tabulary}
\usepackage{nicefrac}		
\usepackage[flushleft]{threeparttable}
\usepackage[ruled,vlined]{algorithm2e}
\usepackage{hyperref}
\SetKwRepeat{Do}{do}{while}
\usepackage{rotating}
\usepackage{tabularx,ragged2e,booktabs}
\usepackage[justification=centering]{caption}
\usepackage{listings}
\usepackage{pifont}
\usepackage[most]{tcolorbox}
\definecolor{backcolour}{rgb}{0.95,0.95,0.91}
\usepackage{lstlinebgrd}
\lstdefinestyle{customc}{
escapechar=?,
  abovecaptionskip=-5pt,
  breaklines=true,
  numbers=left, 
  floatplacement=t,
  frame = TRBL,
  xleftmargin=\parindent,
  language=C,
  showstringspaces=false,
  basicstyle=\footnotesize\ttfamily,
  keywordstyle=\bfseries\color{green!40!black},
  commentstyle=\itshape\color{purple!40!black},
  identifierstyle=\color{blue},
  stringstyle=\color{orange},
}
\newcolumntype{C}[1]{>{\Centering}m{#1}}

\usepackage{mathtools}
\usepackage{verbatim}

\usepackage{soul}
\usepackage{xcolor}

\usepackage{acronym}
\acrodef{FuCE}{fuzzing and concolic execution}
\acrodef{HLS}{high-level synthesis}
\acrodef{IP}{intellectual property}
\acrodef{DUT}{design-under-test}
\acrodef{COTS}{commercial-off-the-shelf}
\acrodef{SoC}{system-on-chip}
\acrodef{CGF}{coverage-guided greybox fuzzing}
\acrodef{AFL}{American fuzzy lop}
\acrodef{S2E}{symbolic executer}
\acrodef{HDL}{hardware description language}
\acrodef{SAT}{Satisfiability}
\acrodef{AFL-SHT}{AFL-SHT}
\acrodef{SCT-HTD}{SCT-HTD}
\acrodef{BCOV}{Branch coverage}
\acrodef{LCOV}{Line coverage}
\acrodef{FCOV}{function coverage}
\acrodef{RTL}{register-transfer level}
\acrodef{IR}{intermediate representation}
\acrodef{SHT}{Synthesizable Hardware Trojan}

\usepackage{pifont}
\newcommand{\cmark}{\ding{51}}%
\newcommand{\xmark}{\ding{55}}%

\hypersetup{
    colorlinks,
    linkcolor={red!50!black},
    citecolor={blue!50!black},
    urlcolor={red!50!black}
}
\usepackage{comment}

\begin{document}

\title[FuCE: Fuzzing+Concolic Testing]{FuCE: \underline{Fu}zzing+\underline{C}oncolic \underline{E}xecution guided Trojan Detection in Synthesizable Hardware Designs}


\author{Mukta Debnath}
\authornotemark[1]
\email{mukta_t@isical.ac.in}
\affiliation{%
  \institution{Indian Statistical Institute}
  \city{Kolkata}
  \country{India}}
  
\author{Animesh Basak Chowdhury}
\authornote{Equal contribution while at Indian Statistical Institute}
\email{abc586@nyu.edu}
\affiliation{%
  \institution{New York University}
  \state{New York}
  \country{USA}
}

\author{Debasri Saha}
\email{debasri_cu@yahoo.in}
\affiliation{%
  \institution{A.K. Chowdhury School of IT,University of Calcutta}
  \city{Kolkata}
  \country{India}
}

\author{Susmita Sur-Kolay}
\email{ssk@isical.ac.in}
\affiliation{%
  \institution{Indian Statistical Institute}
  \city{Kolkata}
  \country{India}}



\begin{abstract}
\input{0_abstract}
\end{abstract}



\maketitle

\input{1_introduction}
\input{2_background}

\input{3_problemAndmotivatingEg}
\input{4_methodology}

\input{5_experimentalResults}
\input{6_conclusion}

\bibliographystyle{ACM-Reference-Format}
\bibliography{sample-base}


\end{document}

%% file: 0_abstract.tex
High-level synthesis (HLS)  is the next emerging trend for designing complex customized architectures for applications such as Machine Learning, Video Processing. It provides a higher level of abstraction and freedom to hardware engineers to perform hardware software co-design. However, it opens up a new gateway to attackers to insert hardware trojans. Such trojans are semantically more meaningful and stealthy, compared to gate-level trojans and therefore are hard-to-detect using state-of-the-art gate-level trojan detection techniques. Although recent works~\cite{fuzzSystemC,symbolicSystemC} have proposed detection mechanisms to uncover such stealthy trojans in \ac{HLS} designs, these techniques are either specially curated for existing trojan benchmarks or may run into scalability issues for large designs. In this work, we leverage the power of greybox fuzzing combined with concolic execution to explore deeper segments of design and uncover stealthy trojans. Experimental results show that our proposed framework is able to automatically detect trojans faster with fewer test cases, while attaining notable branch coverage, without any manual pre-processing analysis.

%% file: 1_introduction.tex
\section{Introduction}
\label{label:intro}

In recent times, hardware designers are increasingly using \ac{COTS} third-party intellectual property (3PIP) for designing complex architectures \cite{rosetta}. For ease of automated design space exploration and functional verification, designers are adopting \ac{HLS} framework like SystemC, synthesizable C/C++. As the designs have become increasingly more complex, chip designers build complex system-on-chips using 3PIPs of required functionalities and integrate them in-house. Post integration, these system-on-chips are outsourced to off-shore fabrication facilities. This typical chip design flow has enabled 3PIP vendors to maliciously inject stealthy bugs at a higher abstraction level that can later be exploited by the attacker to cause malfunction. Thus, verifying the security aspects of such  3PIPs is primarily important for the in-house integrator apart from ensuring correct functionality.

Hardware Trojans at high-level synthesized designs have been recently explored in \cite{s3cbench}, where the attacker injects malicious functionality either by leaking crucial information or corrupt final output of \ac{DUT}. Although there exists a plethora of work to detect hardware trojans at \ac{RTL} or gate-level designs, these detection mechanisms are biased towards certain types of gate-level trojans. The trojans inserted at a higher abstraction level are semantically meaningful and stealthy, but difficult for the defender to precisely detect any abnormal functionality. 
An attacker can manifest a trojan by simply adding an \ac{RTL} statement that uses hardware blocks designed for other functionalities. This motivates performing security testing on high-level synthesized designs. 

In this article, we propose a scalable trojan detection framework based on \ac{FuCE}: combines greybox fuzzing \cite{afl} and concolic execution \cite{sen2007concolic} in a synergistic way to alleviate the downsides of those two approaches in standalone mode. We show that prior state-of-the-art trojan detection works are heavily confined to the type and functionality of trojans, and fail on subtly modified trojan behavior. Our primary contributions in this paper are two-fold:
\begin{enumerate}
    \item a hybrid test generation approach combining the best of both worlds: greybox fuzzing and concolic testing --- our proposed framework complements both the techniques extenuating the problems associated with standalone approaches;
    \item to the best of our knowledge, ours is the first work combining fuzzing with concolic testing to reach deeper segments of \ac{HLS} designs without hitting the scalability bottleneck.
\end{enumerate}

The rest of the paper is organized as follows: Section~\ref{sec:background} outlines the background and the prior related works in the area of trojan detection. Section~\ref{sec:limitAndChallenge} describes the current limitations and challenges using state-of-art techniques. In Section~\ref{sec:framework}, we propose our \ac{FuCE} framework and show the efficacy of results in Section~\ref{sec:results}. Section~\ref{analysis} presents an empirical analysis of the results and concluding remarks appear in Section~\ref{sec:conclusion}.

%% file: 2_background.tex
\section{Background}
\label{sec:background}

\subsection{Overview of Hardware Trojan}
\label{label:background_overview}
For decades, the silicon chip was considered as the root-of-trust of a complex system. The assumption was that the hardware blocks and modules are trustworthy and functions are exactly as specified in the design documentation. However, at the turn of this millennium, researchers have extended the attack surface to the underlying hardware, and showed that it can be tampered to gain privileged information and/or launch denial-of-service attacks. This has disrupted the root-of-trust assumption placed on hardware. The core philosophy of hardware trojan is to insert a malicious logic in the design and bypass the functionality verification of the design. The malicious logic is activated by the attacker's designed input. Since the last decade, hardware trojan design and detection have been extensively studied in the field of hardware security. Security researchers have proposed numerous trojan threat models and novel ways of detecting them. In \cite{decadeTrojan}, the authors have shown hardware trojans can be inserted in various levels of the chip design life-cycle. Thus, security testing of a design becomes extremely important before being passed on to the next stage.



\subsection{Threat model}

\begin{figure}[h]
\centering
\includegraphics[width=0.5\textwidth]{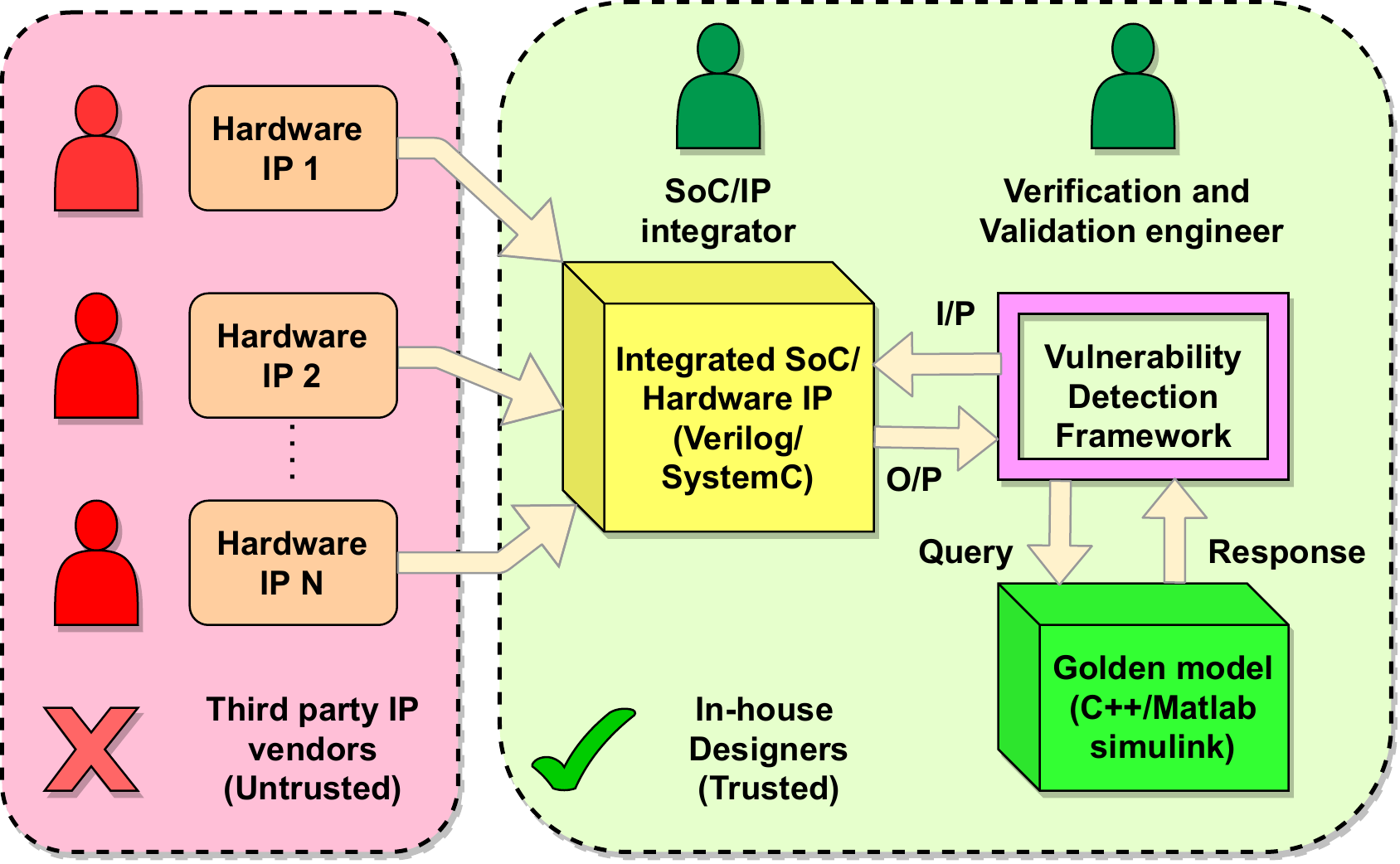}
\caption{Threat model showing untrusted third-party IPs are used for developing customized complex system-on-chip modules}
\label{fig:threatModel}
\end{figure}

We have outlined our threat model in \autoref{fig:threatModel}. We assume in-house designers and engineers are trusted entities who are primarily responsible for developing complex \ac{SoC} modules. A number of third-party hardware IPs are procured from third-party vendors for developing such complex SoC design. However, 3PIPs are untrusted (not developed in-house) and may have hidden backdoors and vulnerabilities. Therefore, it is important for the in-house designers to locate the presence of malicious functionalities in such 3PIPs. For validation purpose, we assume that the in-house designers have access to functionally correct cycle accurate behavioural model of the SoC design (in the form of C++/Simulink model). Our threat model is in line with several earlier works such as~\cite{fuzzSystemC,symbolicSystemC}.

\subsection{Security Testing}
\label{label:background_secTesting}
In the software community, security testing is one of the mandated steps adopted by the practitioners to analyze and predict the behaviour of the system with unforeseen inputs. This has helped develop robust software that are immune against a variety of attacks like buffer-overflow~\cite{bufferOverflow}, divide by zero~\cite{diveByzero}, arithmetic overflow~\cite{arithmeticOverflow}. We describe next two well-known security testing methodologies that are widely used for the same:

\subsubsection{Greybox fuzzing} Fuzz testing is a well known technique in software domain for detecting bugs. Greybox fuzzing~\cite{afl} involves instrumenting the code segments at the point-of-interest, and generate interesting test vectors using evolutionary algorithms. Instrumentation injects markers in the design code which post compilation and execution can track whether a test-case has reached the marker location. A fitness function is used in order to evaluate the quality of a test-vector. Typically, greybox fuzzing is used to improve branch-pair coverage~\cite{afl} of the design, therefore the codes are annotated at every basic block. A test-vector is regarded as interesting, if it reaches a previously unexplored basic block, or hits it for a unique number of times. The fuzz engine maintains a history-table for every basic block covered so far, and retains interesting test vectors for further mutation/crossover. The test generation process completes once the user-defined coverage goal is achieved. There exists a plethora of works \cite{safl,driller,verifuzz} that have improved the performance of greybox fuzzing by augmenting various program analysis techniques. Popular \ac{CGF} engine like \ac{AFL}~\cite{afl} have been able to detect countless hidden vulnerabilities in well-tested softwares.


\subsubsection{Concolic Testing} Concolic testing~\cite{sen2007concolic} is a scalable way of performing symbolic execution on a program with certain inputs considered concrete, and the rest are symbolic inputs. Symbolic execution in general suffers from scalability issues since the number of path constraints generated, are exponential in terms of the number of conditional statements. In order to avoid costly computations, concolic execution executes the program along the path dictated by concrete input and fork execution at branch points. The path constraints generated in concolic execution have reduced the number of clauses and variables, thereby making it easier for solvers and can penetrate deep into complex program checks.  \emph{Driller}~\cite{driller} and \ac{S2E}~\cite{s2e} are examples of engines adopting this approach.


\subsection{Testing for Hardware Trojan detection}

Testing based trojan detection is a well studied problem in the recent past. In earlier works \cite{mero,banga2011odette,saha2015improved,bchowdhury2018,mers}, authors have assumed that the trojan netlist contain a \emph{rare} logic value and/or switching activity at certain nodes. Therefore, the test generation was tuned to excite such nodes in a netlist. Later, researchers  used concolic testing approaches to detect trojans in behavioural level \ac{RTL} designs \cite{iccd2017_concolic,itc2018_concolic,vlsid2018_atpgModelChecking,date2019_concolic}. The objective of performing concolic testing is to penetrate into deeper conditional statements of a HDL program to expose the trojan behaviour. Although the techniques seem to work well on trojan benchmarks \cite{2017_rtlBenchmarks}, employing concolic testing without an efficient search heuristic and a target of interest to cover, results in multiple SAT solver calls taking considerable time for test vector generation. With recent success of coverage-guided greybox fuzzing in software domain, it has been recently adopted in \cite{rfuzz,fuzzSystemC,snpfuzzing} for detecting trojans in hardware design. Concolic test generation for trojan detection in high level designs have only been proposed recently in \cite{2018_concolic_sysC,symbolicSystemC}. We summarize works related to ours in ~\autoref{table:priorWork}.
\begin{table}[!h]
\centering
\caption{Works on Test-based Trojan Detection in Hardware designs}
\resizebox{\textwidth}{!}{
\begin{tabular}{@{}ccccc@{}}
\toprule \toprule
\textbf{Work} & \textbf{Abstraction level} & \textbf{Technique used} & \textbf{Benchmarks} & \textbf{Golden-model available?} \\
\midrule \midrule
\multirow{2}{*}{Chakraborty \textit{et al.}~\cite{mero}} & \multirow{2}{*}{Gate level} & \multirow{2}{*}{Guided ATPG} & \multirow{2}{*}{ISCAS85, ISCAS89} & \multirow{2}{*}{\cmark} \\
\multirow{2}{*}{Banga \textit{et al.}~\cite{banga2011odette}} & \multirow{2}{*}{Gate level} & \multirow{2}{*}{Novel DFT} & \multirow{2}{*}{ISCAS89} & \multirow{2}{*}{\cmark} \\
\multirow{2}{*}{Saha \textit{et al.}~\cite{saha2015improved}} & \multirow{2}{*}{Gate level} & \multirow{2}{*}{Genetic algorithm + SAT formulation} & \multirow{2}{*}{ISCAS85, ISCAS89} & \multirow{2}{*}{\cmark} \\
\multirow{2}{*}{Chowdhury \textit{et al.}~\cite{bchowdhury2018}} & \multirow{2}{*}{Gate level} & \multirow{2}{*}{ATPG binning + SAT formulation} & \multirow{2}{*}{ISCAS85, ISCAS89, ITC99} & \multirow{2}{*}{\cmark} \\
\multirow{2}{*}{Huang \textit{et al.}~\cite{mers}} & \multirow{2}{*}{Gate level} & \multirow{2}{*}{Guided ATPG} & \multirow{2}{*}{ISCAS85, ISCAS89} & \multirow{2}{*}{\cmark} \\
\multirow{2}{*}{Liu \textit{et al.}~\cite{lyu2021maxsense}} & \multirow{2}{*}{Gate level} & \multirow{2}{*}{Genetic algorithm + SMT formulation} & \multirow{2}{*}{TrustHub~\cite{trusthub}} & \multirow{2}{*}{\cmark} \\
\multirow{2}{*}{Ahmed \textit{et al.}~\cite{iccd2017_concolic}} & \multirow{2}{*}{Register transfer level} & \multirow{2}{*}{Concolic testing} & \multirow{2}{*}{TrustHub~\cite{trusthub}} & \multirow{2}{*}{\cmark} \\
\multirow{2}{*}{Ahmed \textit{et al.}~\cite{itc2018_concolic}} & \multirow{2}{*}{Register transfer level} & \multirow{2}{*}{Greedy concolic testing} & \multirow{2}{*}{TrustHub~\cite{trusthub}} & \multirow{2}{*}{\cmark} \\
\multirow{2}{*}{Cruz \textit{et al.}~\cite{vlsid2018_atpgModelChecking}} & \multirow{2}{*}{Register transfer level} & \multirow{2}{*}{ATPG + Model checking} & \multirow{2}{*}{TrustHub~\cite{trusthub}} & \multirow{2}{*}{\cmark} \\
\multirow{2}{*}{Liu \textit{et al.}~\cite{date2019_concolic}} & \multirow{2}{*}{Register transfer level} & \multirow{2}{*}{Parallelism + concolic testing} & \multirow{2}{*}{TrustHub~\cite{trusthub}} & \multirow{2}{*}{\cmark} \\
\multirow{2}{*}{Pan \textit{et al.}~\cite{pan2021automated}} & \multirow{2}{*}{Register transfer level} & \multirow{2}{*}{Reinforcement learning} & \multirow{2}{*}{TrustHub~\cite{trusthub}} & \multirow{2}{*}{\cmark} \\
\multirow{2}{*}{Le \textit{et al.}~\cite{fuzzSystemC}} & \multirow{2}{*}{HLS/SystemC} & \multirow{2}{*}{Guided greybox fuzzing} & \multirow{2}{*}{S3C~\cite{s3cbench}} & \multirow{2}{*}{\cmark} \\
\multirow{2}{*}{Bin \textit{et al.}~\cite{symbolicSystemC}} & \multirow{2}{*}{HLS/SystemC} & \multirow{2}{*}{Selective symbolic execution} & \multirow{2}{*}{S3C~\cite{s3cbench}} & \multirow{2}{*}{\cmark} \\
\multirow{2}{*}{\textbf{\textit{FuCE (Ours)}}} & \multirow{2}{*}{HLS/SystemC} & \multirow{2}{*}{Greybox fuzzing + Concolic Execution} & \multirow{2}{*}{S3C~\cite{s3cbench}} & \multirow{2}{*}{\cmark} \\
&&&& \\ \bottomrule
\end{tabular}
}
\label{table:priorWork}
\end{table}
\subsection{Trojan detection in high-level design}
With high level synthesis becoming the new trend for designing customized hardware accelerators, only few works~\cite{s3cbench,farimah2021} have studied hardware trojan and security vulnerabilities in \ac{HLS} designs and proposed preliminary countermeasures to detect them. In prior work\cite{fuzzSystemC}, Trojan inserted in high-level design is called synthesizable hardware Trojan (SHT) as it gets manifested as malicious backdoor in the hardware design. Therefore, the problem of Trojan detection in low level RTL design can be appropriately abstracted as finding SHT in high-level design. Till date, \cite{fuzzSystemC} and \cite{symbolicSystemC} have systematically addressed the problem of Trojan detection in HLS designs. In \cite{fuzzSystemC}, the authors have tuned the software fuzzer \ac{AFL}~\cite{afl} for S3C Trojan benchmark characteristics and showed that their technique outperforms the vanilla \ac{AFL}. The modified \ac{AFL} called \ac{AFL-SHT}, introduces program-aware mutation strategy to generate meaningful test vectors. In \cite{symbolicSystemC}, the authors identify the additional overhead of concolic testing from usage of software libraries, and have restricted the search space within the conditional statements of design. They named their automated prototype as \ac{SCT-HTD}, based on \ac{S2E}~\cite{s2e}. In the next section, we focus on studying the Trojan characteristic embedded in high-level synthesized design and evaluate the efficacy of existing detection techniques.

%% file: 3_problemAndmotivatingEg.tex
\section{Limitations and challenges}
\label{sec:limitAndChallenge}
We present our motivating case-study on a real-time finite state machine implementation. We use a system controller design mimicking a typical hardware functionality and highlight limitations of existing techniques to discover the trojan behavior.

\subsection{Motivating case-study}

\begin{lstlisting}[caption=Motivating example,style=customc,linebackgroundcolor={\ifnum\value{lstnumber}=19\color{yellow}\fi\ifnum\value{lstnumber}=15\color{yellow}\fi\ifnum\value{lstnumber}=16\color{yellow}\fi\ifnum\value{lstnumber}=17\color{yellow}
    \fi\ifnum\value{lstnumber}=18\color{yellow}
    \fi},label={label:motivatingCode}]
int controller(int stateA, int stateB)
{
  unsigned long long cycle=0,input = 101 ;
  bool switchA = FALSE;
    
  stateA+=input ; stateB-=input ;
  if(stateA != 23978 || stateB != 5687) ?\label{line:stateCheck}?
    exit(0) ;

  while(input){
      if(stateA == 23978 && stateB == 5678 && switchA == FALSE) ?\label{line:cond1}? // Branch 1
      {
        switchA = TRUE ; cycle++ ;
        swap(stateA,stateB) ;
        if(cycle >= 2^20-1) ?\label{line:trojan}? //Trojan 
        {
            switchA = FALSE ;
            swap(stateA,stateB) ;
        }
      }
      else if(switchA == TRUE) ?\label{line:cond2}? // Branch 2
      {
        stateA += 100 ; stateB -= 100 ;
        if(stateA == 23978 && stateB == 5678) ?\label{line:cond3}? // Branch 3
            switchA = FALSE ;
      }
      scanf("\n%d",input) ;
  }
}
\end{lstlisting}
    
In~\autoref{label:motivatingCode}, we have a code-snippet of a typical controller accepting state information \texttt{stateA} and \texttt{stateB}. The controller first checks if the \texttt{stateA} and \texttt{stateB} are set to the values $23978$ and $5678$, respectively (line \autoref{line:stateCheck}). Once the guard condition is satisfied, it enters a while loop, and check for the values of \texttt{stateA}, \texttt{stateB} and \texttt{switchA}. The loop traversed for the first time, satisfies $Branch~ 1$ (line \ref{line:cond1}) and swaps the values of \texttt{stateA} and \texttt{stateB}, setting \texttt{switchA} to TRUE. In subsequent iterations, $Branch~ 2$ is always satisfied as \texttt{switchA} = TRUE (line~\ref{line:cond2}), resulting in updating the values of \texttt{stateA} and \texttt{stateB}. It also checks whether \texttt{stateA} and \texttt{stateB} have reached the pre-defined values (line \ref{line:cond3}). If so, then \texttt{switchA} is set to FALSE. At the end, the controller accepts an input from the user for performing further action. 

\begin{figure}[t]
\centering
\includegraphics{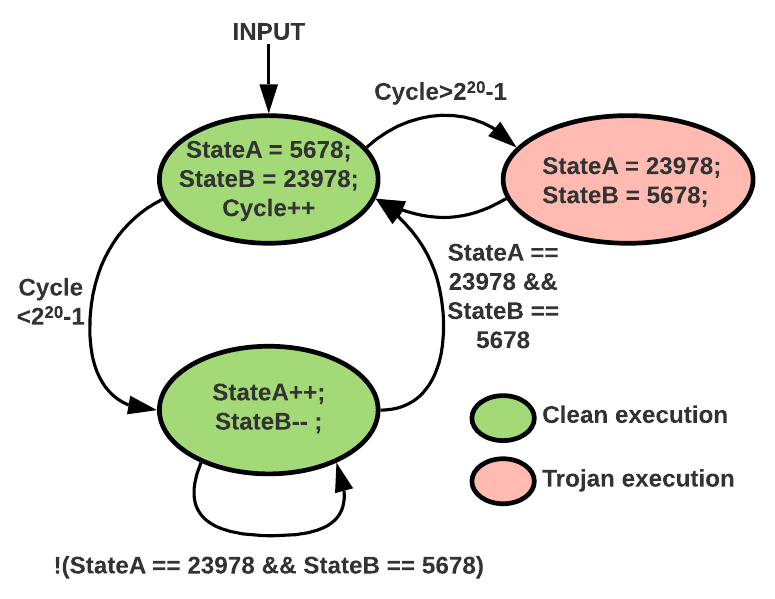}
\caption{State-transition diagram of motivating example~[\ref{label:motivatingCode}]}
\label{fig:state_diag}
\end{figure}

We insert a Trojan code at line \ref{line:trojan} where we compare the current value of \texttt{cycle} with a very large number. The attacker intends to skip the functionality of {\it Branch 2}, and provide a false impression to the user that the controller is working by accepting inputs, however skipping {\it Branch 2} and  not performing any operations. We explain it with the help of a state-transition diagram in ~\autoref{fig:state_diag}. Here, we observe that as soon as the \texttt{cycle} reaches the value of $2^{20}-1$, \texttt{switchA} is maliciously set to FALSE and the values of \texttt{stateA} and \texttt{stateB} are swapped instead of the expected updates in {\it Branch 2}. Thus, the loop body repeatedly executes {\it Branch 1} and maliciously increments \texttt{cycle} after accepting input from the user. The Trojan resembles closely to the ``ticking time-bomb" or ``sequential Trojan" behaviour where the system malfunctions after running for sufficiently large number of cycles. From test and verification perspective, simulating the design for large number of cycles in an exhaustive manner is a hard problem.

\subsection{Evaluating AFL-SHT on this example}
\label{subsec:afl-sht}
Le \textit{et al.}\cite{fuzzSystemC} envisioned hardware trojans in high-level SystemC designs as \ac{SHT}. They proposed fuzzing-based test generation \ac{AFL-SHT} using \ac{AFL} as the backend engine. Initially, the authors evaluated the trojan detection capability of vanilla \ac{AFL} on \textit{S3C} benchmarks and identified the pitfalls of \ac{AFL} in detecting \ac{SHT}. They presented three major modifications in the mutation block of fuzz-engine: 1) pump mutation 2) format-aware trimming, and 3) design-aware \textit{interesting number} generation to tune test-generation for trojan detection.

We evaluated our motivating example with the authors' version of \ac{AFL-SHT} and found that it was unable to generate appropriate values of \texttt{input}, \texttt{stateA} and \texttt{stateB}  to activate the trojan behaviour. We observe that in-spite of aiding the fuzz-engine with coarse-grain information about the \textit{interesting numbers}, it was unable to explore beyond line~\ref{line:stateCheck}. This clearly shows the incapability of customized fuzz engine to generate \texttt{stateA}$ =23978$ and \texttt{stateB}$ = 5678$ and satisfy the branch constraints. Similarly, we modified the trigger conditions of trojans in \textit{S3C} benchmarks and found that \ac{AFL-SHT} was unable to detect trojans while the performance were slightly better than vanilla \ac{AFL}. This indicates \ac{AFL-SHT} is unable to utilize the pump mutation and using interesting numbers from benchmarks in an effective manner. We conclude our evaluation of \ac{AFL-SHT} with a simple takeaway message: fuzzing needs additional aid to explore code segments guarded by complex conditional checks.


\subsection{Evaluating SCT-HTD on this example}
A growing body of work in high-level synthesized designs have led to renewed interest in concolic testing for SystemC~\cite{2018_concolic_sysC,symbolicSystemC,2016_concolic}. Recent work \ac{SCT-HTD}~\cite{symbolicSystemC} has proposed a scalable, selective and systematic exploration approach for concolic testing of SystemC designs to uncover stealthy trojans. The authors identified a crucial insight that concolic engine do not distinguish in-built between library codes and design codes, and therefore gets stuck in exploration of undesirable library codes. The underlying assumption is: library codes and pragmas are maintained by SystemC specifications and therefore are trusted elements. Thus, exploring different paths in library codes does not hold much relevance and hence the authors have selectively restricted the state-space exploration within the design code. Additionally, while exploring the state-space, they prioritize states hitting uncovered conditions compared to states having already explored conditions. The authors have evaluated their approach on \textit{S3C} benchmarks and showed improvement in terms of number of inputs generated for trojan detection.

A direct advantage of using concolic based approach for test-generation is systematic exploration of state-space. It maintains a history of states previously visited and prioritize the bandwidth towards exploring complex conditional checks. We run selective concolic engine (re-implementing the concepts of \ac{SCT-HTD}) and run it on our motivating example. We discovered that \ac{SCT-HTD} failed to trigger the Trojan condition particularly because of two reasons: 1) \ac{SCT-HTD} forked two states for every iteration of the while loop and soon running into out of memory on a machine having 16 GB RAM, 2) \ac{SCT-HTD} repeatedly invokes SAT-engine to generate test-input for satisfying the condition at line~\ref{line:cond1} when the condition at line~\ref{line:cond2} is True. On a high-level, it shows that concolic engines require an additional aid for intelligent exploration of search space leaving less memory footprint.




\subsection{Lessons learnt}
Post evaluation of state-of-art techniques of trojan detection, we conclude that the challenges lying ahead involves two important issues: 1) detecting trojans in a faster and scalable manner, and 2) detecting extremely "hard-to-trigger" trojan logic. As a defender, one can never make a prior assumption about possible location of trojans and tune the Testing methodology in a particular way. For a defender, the confidence of a design being ``Trojan-free" can only come when a test-set is generated providing a sufficient coverage on the design. Our evaluation with fuzzing and concolic testing shows that they can be combined in a synergistic way to accelerate trojan detection. This can avoid the path explosion problem of symbolic execution by controlled forking in symbolic loops using the fuzzer-generated test cases, thereby reaching deeper code segments without generating lots of states.






\begin{figure}[h]
\begin{center}

\includegraphics[width=\textwidth]{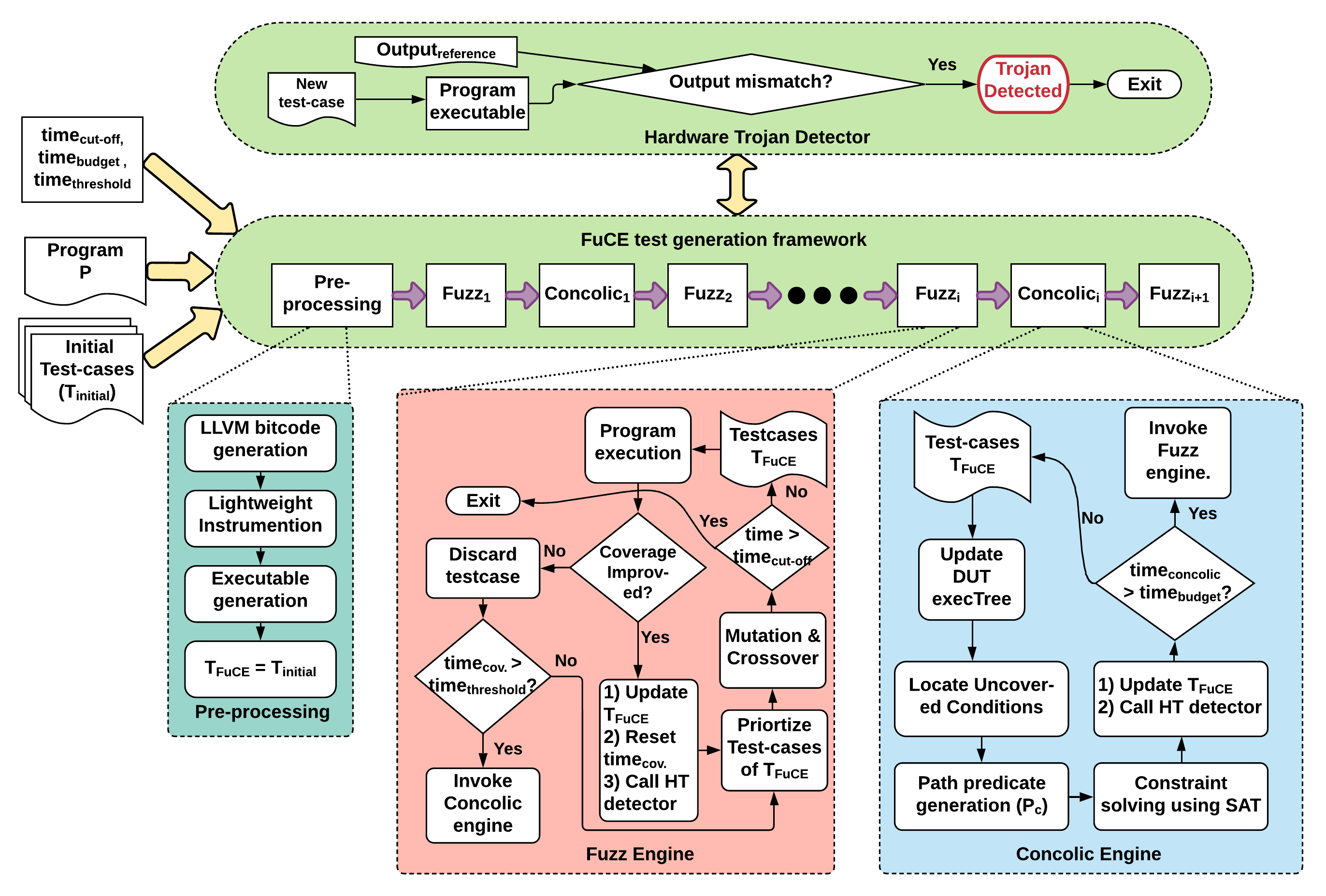}
\end{center}
\caption{\ac{FuCE} test generation framework. \emph{Fuzz engine} is fed with initial test-cases. As coverage improvement ceases in fuzz engine, \emph{Concolic engine} starts execution with fuzz generated test cases. \emph{Fuzz engine} and \emph{Concolic engine} execute sequentially to penetrate deep into hard-to-satisfy conditional checks of programs. The \emph{Trojan Detector} checks for trojan whenever \emph{FuCE} generates a new test-case.}

\label{fig:flow_chrt}
\end{figure}
Combining the two mainstream techniques for security testing, FuCE can achieve high code coverage with faster Trojan detection ability. FuCE avoids getting stuck either in Fuzz testing or in symbolic execution. FuCE can avoid the path explosion problem of symbolic execution to a certain extend, by controlled forking in symbolic loops using the fuzzer-generated test cases.  So it can reach deeper code areas without generating lots of states.
{\small
\begin{algorithm}[!ht]
\SetAlgoLined
\KwData{Design Under Test $DUT$, User provided test-inputs $T_{initial}$, User defined time bound $time_{cut-off}$} 


\KwResult{{$T_{fuzzed}$}       
   \Comment{Interesting test-inputs queue}}
$T_{fuzzed} \gets T_{initial}$ \Comment{Initialization of the AFL's test-inputs queue}\\ 
\While{time $ \leq time_{cut-off}$}{
    \For{$\tau \in T_{fuzzed}$} 
   {
        \Comment{Mutate $\tau$ to generate test-cases based on the energy parameter}\\
        $K\gets$ CALCULATE\_ENERGY($\tau$)\\
        \For{$i \in \{1,2,\dots,K\}$}{
            $\tau'\gets$ MUTATE-SEED($\tau$) \Comment{$\tau'$ denotes the mutated test case} \\
            \If{IS-INTERESTING($DUT, \tau'$)}
            {
            $T_{fuzzed}\gets$ $T_{fuzzed}\cup \tau'$
            \Comment{$\tau'$ is interesting if it improves branch coverage}\\}
        }
        }
    }
    \Return $T_{fuzzed}$
\caption{FUZZER($DUT$, $T_{initial}$)}
\label{algo:AFL_FUZZ}
\end{algorithm}
}

%% file: 4_methodology.tex
\section{FuCE framework}
\label{sec:framework}
We present the workflow of \ac{FuCE} in~\autoref{fig:flow_chrt}. The \ac{FuCE} framework consists of three components: 1) Greybox fuzzing, 2) Concolic execution, and 3) Trojan detector.


\subsection{Greybox Fuzzing by AFL}
High-level synthesized designs predominantly written in SystemC/C++ are initially passed through static analysis tool \texttt{LLVM}~\cite{LLVM} backend to generate \ac{IR}. The \texttt{LLVM} generated \acp{IR} are fed to \texttt{afl-clang-fast}, which is based on \texttt{clang}~\cite{clang}, a front end compiler for programming languages like C, C++, SystemC, among others.


\texttt{afl-clang-fast} performs code instrumentation by automated injection of control flow statements on every conditional statement in run time, and generates executable. The core insight is: Trojan logic must be embedded under one of the conditional statements, so covering all conditional statements while verification and testing provides sufficient confidence of triggering the Trojan logic. The instrumented executable is then fed to our greybox fuzz-engine \ac{AFL} along with an initial test-set ($T_{initial}$) for fuzz testing.



In~\autoref{algo:AFL_FUZZ}, we outline the overall flow of Greybox fuzzing. At first, we provide the high-level \ac{DUT} and a user-provided test-set $T_{initial}$ to the fuzzing framework. 
The \textit{CALCULATE-ENERGY} function assigns energy to the initial seed $T_{initial}$ on the basis of external features of the test case like the execution time, bitmap coverage, depth of the test case in terms of fuzzing hierarchy. A test case that is fast, covers more branches and has more depth, is given more energy. AFL then decides the number of random fuzzing iterations for that test case. \ac{AFL} uses $T_{initial}$ to perform operations like \textit{deterministic mutations} and \textit{havoc mutations} to generate newer test-cases with the help of the \textit{MUTATE-SEED} function. The deterministic mutation stage scans each byte of test-case and mutates them to generate new test-case. This includes bit flipping, byte flipping, arithmetic increments and decrements and magic value substitution. The number of children test-cases generated loosely depends on the size of the original test-case. However, havoc mutation performs aggressive mutations like mutating bit/bytes values at random location with a random value, deleting or cloning sub-sequence for generating new test-cases. \ac{AFL} uses branch-pair as a fitness metric to determine the quality of test-input. For each branch-pair, \ac{AFL} maintains a hash-table entry the number of times it is hit. The \textit{IS-INTERESTING} function checks whether the mutated test case is interesting or not. \ac{AFL} considers a test-input to be \textit{interesting}, if it covers a new-branch pair not hit so far, or has hit a branch-pair unique number of times compared to past observations. Interesting test-inputs are retained to form the next candidates for fuzzing. The algorithm terminates when either no more interesting test cases can be found or the user-defined $time_{cut-off}$ expires. \ac{AFL} maintains all interesting test-inputs in the queue $T_{fuzzed}$.



\subsection{Concolic Execution by S2E}
\label{label:s2e_engine}
In our work, we use \ac{S2E} as our concolic execution engine for test-generation. \ac{S2E} has two main components: 1) a concolic virtual machine based on QEMU~\cite{qemu} and 2) a symbolic execution engine based on KLEE to switch back and forth between concrete execution and symbolic execution. We provide it the high level design \ac{DUT} and a set of test-cases $T_{initial}$. The \textit{CONC-EXEC} execute the \ac{DUT} with all test-cases $T_{initial}$ generating concrete execution traces. \ac{S2E} maintains an execution tree $DUT_{execTree}$ and identifies all the true and/or false edges of conditional nodes which are not covered by $T_{initial}$. \ac{S2E} then assigns symbolic values to those predicates. The \textit{COND-PREDICATE} constructs the path constraints for the uncovered edge of a condition, forks a new thread and invokes SAT-solver (\textit{CONSTRAINT-SOLVER}) to generate the test-case. \ac{S2E} selects the path for exploration in depth-first search order, based on its coverage analyzer and heuristically selects the path that maximizes the coverage. So, the final test-cases reported by \ac{S2E} ideally should cover all conditions of $DUT_{execTree}$. We outline this approach in~\autoref{algo:concolic_execution}.

{\small
\begin{algorithm}[t]
\SetAlgoLined
\KwData{Design Under Test $DUT$, User provided test-inputs $T_{initial}$}
\KwResult{{$T_{concolic}$} \Comment{Set of test-cases generated by the concolic engine}} 
$DUT_{execTree} \gets \phi$ \Comment{Execution tree for DUT}\\
\For{$\tau \in T_{initial}$}
{
    \Comment{Update DUT's execution tree with path traces obtained from concrete execution of initial test inputs}
    $P_{trace} \gets $ CONC-EXEC($DUT$,$\tau$)\\
    $DUT_{execTree} \gets DUT_{execTree} \cup P_{trace}$
}
\For{uncovered cond $c \in DUT_{execTree}$}
{
    \Comment{Perform symbolic execution steps targeting uncovered conditional statements}\\
   $p_{c} \gets$ COND-PREDICATE($c$) \\
   $t_{new}$ $\gets$ CONSTRAINT-SOLVER($p_c$) \Comment{$t_{new}$ is newly generated test-case by the concolic engine}\\
    $T_{concolic}$ $\gets$ $T_{concolic}$ $\cup$ $t_{new}$
    
} 
\Return $T_{concolic}$
\caption{CONCOL-EXEC($DUT$,$T_{initial}$)}
\label{algo:concolic_execution}
\end{algorithm}
}

\begin{algorithm}[!ht]
\SetAlgoLined
\KwData{Design Under Test ($DUT$), Initial test-inputs ($T_{initial}$), Time limit ($time_{cut-off}$), Threshold time limit for coverage improvement by Fuzzer ($time_{threshold}$), Time budget allocated for concolic execution ($time_{budget}$)}
\KwResult{{$T_{FuCE}$} \Comment{Final test-cases generated by \ac{FuCE}} }
\Comment{\textit{InvokeConcolic} performs concolic execution with fuzzing generated test-inputs}\\
 \SetKwFunction{FInvokeConcolic}{InvokeConcolic} 
  \SetKwProg{Fn}{Function}{:}{}
  \Fn{\FInvokeConcolic{$DUT_{execTree}$,$i$}}{
    $time_{concolic} \gets 0$ \Comment{Monitor concolic execution runtime} \\ 
     \For{uncovered cond $c \in DUT_{execTree}$}
     {
         $p_{c} \gets$ COND-PREDICATE($c$) \\
         $t_{new}$ $\gets$ CONSTRAINT-SOLVER($p_c$) \Comment{$t_{new}$ is a newly generated test case}\\
        $T_{FuCE}$ $\gets$ $T_{FuCE}$ $\cup$ $t_{new}$\\
         Invoke Trojan detector with $t_{new}$ (\autoref{algo:htdetector}) \\
         \If{$time_{concolic} > time_{budget}$}{
            \textbf{break}
         }
     } 
  }
$i \gets 1$                      \Comment{Phase ID of sequential execution}\\
$T_{FuCE} \gets T_{initial}$ \Comment{\ac{FuCE} gets initial test-cases  $T_{initial}$}\\
$time_{coverage} \gets 0$ \Comment{$time_{coverage}$ monitors time elapsed since last test-case retained}\\

\Comment{$time$ denotes wall time of \ac{FuCE} }\\
\While{time $\leq time_{cut-off}$}{
        \For{$\tau \in T_{FuCE}$}
        {
        $K\gets$ CALCULATE\_ENERGY($\tau$) \Comment{Mutate $\tau$ to generate test-cases based on the energy parameter}\\ 
        \For{$j \in \{1,2,\dots,K\}$}{
            $\tau'\gets$ MUTATE-SEED($\tau$) \Comment{$\tau'$ denotes the mutated test case}\\
            \If{IS-INTERESTING($DUT, \tau'$)}
            {
                $T_{FuCE}\gets$ $T_{FuCE}\cup \tau'$\\
                Invoke Trojan detector  with $\tau'$ (\autoref{algo:htdetector})\\
                Reset $time_{coverage}$\\
                }
                \If{$time_{coverage} > time_{threshold}$}
                {
                InvokeConcolic($DUT_{execTree}$, $i$) \Comment{Invokes the concolic engine when the fuzzer gets stuck} ($DUT_{execTree}$ is the program execution tree of $DUT$) \\
                \If{time $> time_{cut-off}$}
                {
                    \Return $T_{FuCE}$ \Comment{User defined total time budget for FuCE is  exhausted}
                }
                \Else{
                    $i = i+1$\\
                    \textbf{break}
                }
            }
            
            }
           
    }
}
$coverage_{FuCE} = \texttt{reportCoverage}(T_{FuCE}$) \Comment{Reports code coverage}
\\
\Return ($T_{FuCE}, coverage_{FuCE}$)
\caption{FuCE ($DUT$, $T_{initial}$, $time_{cut-off}$, $time_{threshold}$, $time_{budget}$)}
\vspace{0.3cm}
\label{algo:FuCE_1}
\end{algorithm}

\subsection{Fusing fuzzer with concolic execution (FuCE)}

We leverage the power of concolic execution to alleviate the drawbacks of greybox fuzzing without hitting a scalability bottleneck. As shown in ~\autoref{fig:flow_chrt}, we first perform lightweight instrumentation on all conditional statements of \ac{DUT} and generate an instrumented executable. We start our fuzz-engine (\textit{FUZZER}) with a set of initial test-cases $T_{initial}$. The fuzz-engine  generates interesting test-cases using genetic algorithm and explores various paths in the design. When there is no coverage improvement and a user-defined time period $time_{threshold}$ is over, the concolic engine (\textit{CONCOL-EXEC}) is invoked for unseen path exploration. The concrete execution function of the concolic engine generates the $DUT_{execTree}$ for the DUT by feeding the fuzzer generated test cases. The \textit{CONCOL-EXEC} identifies uncovered conditions in $DUT_{execTree}$ and forks new threads for symbolic execution on such conditions using depth-first search strategy. We use fuzzed test-cases for concolic engine to generate new test-cases satisfying complex conditional statements. In order to avoid scalability bottleneck,  we limit the runtime of concolic engine to $time_{budget}$. Concolic execution generated test-cases are then fed back to the fuzzer, thereby allowing scalable exploration of deeper program segments. This process continues till trojan is detected, which is the main objective of \ac{FuCE}. Whenever a new test case is generated with either the fuzz engine or the concolic engine, it is added to $T_{FuCE}$, the test case queue for \ac{FuCE},  and the trojan detector is invoked with the latest test case added to $T_{FuCE}$. If trojan is detected successfully, \ac{FuCE} displays the result for trojan detection and stops. We formally present our test-generation approach in \autoref{algo:FuCE_1}.

There may be designs where trojans get detected before 100\% branch coverage is obtained. For related goals of either generating test-cases for achieving 100\% branch coverage or trojan dertection in the absence of a golden model, 
 we can run \ac{FuCE} for a pre-defined $time_{cut-off}$  using a similar flow of switching between the fuzzer and the concolic engine as in~\autoref{fig:flow_chrt}. Instead of checking for trojan detection, we check for complete branch coverage for the design and stop \ac{FuCE} on attaining 100\% branch coverage or on time out.

\begin{algorithm}[h]
\SetAlgoLined
\KwData{$DUT$ is Device Under Test, $Out_{golden}$ is golden model output), $t$ is the new test-case of \ac{FuCE}}
        $out_{ref} \gets Out_{golden}(t)$ \\
        $out_{DUT} \gets SIMULATOR(DUT,t)$ \\
        \If{$out_{DUT} \neq out_{ref}$ }{
            Trojan detected\\
            \texttt{break}
        }

$coverage_{FuCE} = \texttt{reportCoverage}(T_{FuCE}$) \Comment{Reports code coverage}
\\
\Return ($T_{FuCE}, coverage_{FuCE}$)
\caption{HT-DETECTOR ($DUT$, $Out_{golden}, t$)}
\label{algo:htdetector}
\end{algorithm}

\subsection{Hardware Trojan Detection}
\label{label:trojan_detector}
Using \ac{FuCE} test generation framework, we localize trojans in a given $DUT$ (\autoref{algo:htdetector}). For every test-case \ac{FuCE} generated either by \emph{FUZZER} or \emph{CONCOL-EXEC}, we run our design, collect the output response and compare it with reference response. The \textit{SIMULATOR} invokes the SystemC library that provides predefined structures and simulation kernel for simulation of SystemC designs. The assumption is: any deviation of $DUT$'s output (eg., bit corruption at certain locations or report of internal state information) from expected response is considered to be suspicious in nature and triggered from possible trojan behaviour. \ac{FuCE} terminates as soon as a trojan behaviour is detected, otherwise at a user-defined time-out $time_{cut-off}$  and reports coverage metrics using the test-cases generated. 

\subsection{Evaluating FuCE on our motivating example}
We evaluated \ac{FuCE} on our motivating example to check the efficacy of our proposed framework. Our fuzzer quickly generated a considerable number of test-cases for \texttt{stateA} and \texttt{stateB} but was unable to explore beyond line~\ref{line:stateCheck}. The test-cases generated by fuzzer were passed on to concolic engine. Concolic engine generated values of \texttt{stateA} and \texttt{stateB} satisfying the condition at line~\ref{line:stateCheck}. These newly generated test-cases were fed back to the fuzzer leading to faster exploration of the entire loop body. We observed that the fuzzer preserved the test-cases which covered the loop body a unique number of times. Finally, \ac{FuCE} reached the Trojan location within 560s$\pm 15\%$, whereas neither \ac{AFL-SHT} nor \ac{SCT-HTD} could detect it within two hours of run.

%% file: 5_experimentalResults.tex
\section{Experimental setup and design}
\label{sec:results}

\subsection{Experimental setup}

We implement \ac{FuCE} using state-of-art software testing tools: \ac{AFL} (v2.52b)~\cite{afl} for greybox fuzzing and \ac{S2E}~\cite{s2e} to perform concolic execution. For robust coverage measurements, we cross-validated our results using a combination of coverage measuring tools: namely \emph{afl-cov-0.6.1}~\cite{afl-cov}, \emph{lcov-1.13}~\cite{Lcov}, and \emph{gcov-7.5.0}~\cite{Gcov}. Experiments are performed on 64-bit linux machine having $i5$ processor and 16 GB RAM clocked at 3.20 GHz.

\subsection{Benchmark characteristics}
We evaluated \ac{FuCE} on the SystemC benchmark suite, $S3CBench$~\cite{s3cbench_benchmark} having SystemC Trojan-infected designs.
The Trojans have a wide spectrum of purpose ranging across: Denial-of-service, Information leakage, and corrupted functionality. The types of trojan based on their triggered mechanism are categorized in~\autoref{tab:trojan_type}. For the trojan type where the payload has memory, the trojan remains active for a prolonged period of time even when the trigger condition is not active anymore. We define the severity level based on this characteristic. The $S3CBench$ is synthesizable to RTL using any commercial High Level Synthesis (HLS) tool. ~\autoref{tab:bench_char} shows the characterization of the \textit{S3C} benchmark both for the original circuit, and the trojan induced circuit. 


The benchmarks considered have diverse characterization: 1) Image and signal processing: ADPCM, 2) Cryptography: AES, 3) Data manipulation: Bubble sort, 4) Filters: Decimation, and 5) IP protocols: UART. The trojans in S3CBench are hard to detect with random seeds~\cite{s3cbench}. The benchmark suite provides certain test-cases having high statement coverage but does not trigger the trojan behaviour. Two benchmarks in S3C suite, namely \textit{sobel} and \textit{disparity} that accept image as file input, were not considered because the concolic engine was unable to generate test-cases with these input formats. However, we present our results on the largest benchmark \emph{AES-cwom}, and on the most complex benchmark of the $S3C$ suite, i.e.,\emph{interpolation-cwom}, which indeed demonstrates the efficacy of our approach. 


\begin{table}[t]
\centering
\caption{Trojan types --- Combinational: Comb., Sequential: Seq.}
\begin{tabular}{cccc}
\toprule
Trojan & Trigger & Payload & Severity\\
\midrule
\rt{CWOM}   & \rt{Comb.} & \rt{Comb.}   & \rt{Low}      \\ 
\bt{CWM}    & \bt{Comb.} & \bt{Seq.} & \bt{High}     \\ 
\mat{SWOM}   & \mat{Seq.}    & \mat{Comb.}   & \mat{Low}      \\ 
\ct{SWM}    & \ct{Seq.}    & \ct{Seq.} & \ct{High}     \\
\bottomrule
\end{tabular}
\label{tab:trojan_type}
\end{table}

\subsection{Design of experiments}

We perform two variants of experiments to evaluate the efficacy of \emph{FuCE}: 1) Trojan detection 2) Achievable branch coverage. We compare the results with standardized baseline techniques namely fuzz-testing based approach(AFL) and symbolic model checking(S2E). We now describe the experimental setup of each baseline:

\textbf{Baseline 1~(Fuzz testing)}: We run \ac{AFL} on S3C benchmarks using default algorithmic setting. Initial seed inputs are randomly generated.


\textbf{Baseline 2~(Symbolic execution)}: Like baseline 1, we run \ac{S2E} on S3C benchmarks having default configurations. Randomly generated seed inputs are provided to \ac{S2E} as inputs.

\begin{table}[t]
    \centering
    \caption{HLS synthesized hardware characterization of S3C benchmarks. Trojan types: \rt{CWOM}, \ct{SWM} and \mat{SWOM}}
    \begin{tabular}{ccccccccc}
        \toprule
        \multirow{2}{*}{Benchmark} &
        \multirow{2}{*}{Type} &
        \multicolumn{3}{c}{SystemC characterization} &
        \multicolumn{4}{c}{HLS synthesized hardware} \\
        \cmidrule(lr){3-5}
        \cmidrule(lr){6-9}
        & & {Branches} & {Lines} & {Functions} & {LUTs} & {Registers} & {Nets} & {Critical path(ns)}
        \\ 
        \midrule
        \multirow{3}{*}{ADPCM} & orig & 26 & 186 & 6 & 121 & 87 & 346 & 3.94
        \\ & \ct{SWM} & \ct{28} & \ct{186} & \ct{6} & \ct{120} & \ct{118} & \ct{394} & \ct{3.801}
        \\ & \mat{SWOM} & \mat{30} & \mat{187} & \mat{6} & \mat{163} & \mat{240} & \mat{588} & \mat{3.019}
        \\
        \cmidrule(lr){1-9}
        \multirow{2}{*}{AES} & orig & 50 & 371 & 13 & 2782 & 4684 & 8809 & 7.589
        \\ & \rt{CWOM} & \rt{68} & \rt{380} & \rt{13} & \rt{2886} & \rt{4772} & \rt{9039} & \rt{7.668}
        \\
        \cmidrule(lr){1-9}
        \multirow{3}{*}{Bubble\_sort} & orig & 20 & 78 & 3 & 472 & 551 & 1219 & 8.944
        \\ & \rt{CWOM} & \rt{22} & \rt{78} & \rt{3} & \rt{494} & \rt{551} & \rt{1219} & \rt{7.527}
        \\ & \ct{SWM} & \ct{22} & \ct{78} & \ct{3} & \ct{546} & \ct{584} & \ct{1400} & \ct{7.87} \\
        \cmidrule(lr){1-9}
        \multirow{2}{*}{Filter\_FIR} & orig & 14 & 75 & 2 & 68 & 36 & 146 & 5.729
        \\ & \rt{CWOM} & \rt{16} & \rt{75} & \rt{4} & \rt{89} & \rt{59} & \rt{213} & \rt{7.46} \\
       \cmidrule(lr){1-9}
        Interpolation & orig & 10 & 108 & 3 & 984 & 654 & 212 & 7.45
        \\ & \rt{CWOM} & \rt{30} & \rt{108} & \rt{3} & \rt{1071} & \rt{595} & \rt{212} & \rt{8.331}
        \\ & \ct{SWM} & \ct{30} & \ct{108} & \ct{3} & \ct{612} & \ct{570} & \ct{212} & \ct{8.321} 
        \\ & \mat{SWOM} & \mat{30} & \mat{109} & \mat{3} & \mat{612} & \mat{569} & \mat{212} & \mat{8.321}
        \\
        \cmidrule(lr){1-9}
        Decimation & orig & 88 & 304 & 3 & 3018 & 1696 & 634 & 8.702
         \\ & \ct{SWM} & \ct{94} & \ct{304} & \ct{3} & \ct{3108} & \ct{1741} & \ct{634} & \ct{8.702} \\
         \cmidrule(lr){1-9}
        Kasumi & orig & 36 & 288 & 12 & 1378 & 956 & 188 & 8.016
         \\ & \ct{SWM} & \ct{38} & \ct{288} & \ct{12} & \ct{1385} & \ct{958} & \ct{272} & \ct{8.016} \\
         & \rt{CWOM} & \rt{38} & \rt{288} & \rt{12} & \rt{1431} & \rt{987} & \rt{273} & \rt{9.266} \\
        \cmidrule(lr){1-9}
        UART & orig & 28 & 160 & 3 & 510 & 142 & 1336 & 3.137
         \\ & \ct{SWM-1} & \ct{48} & \ct{164} & \ct{3} & \ct{549} & \ct{196} & \ct{1336} & \ct{2.766} 
         \\ & \ct{SWM-2} & \ct{50} & \ct{164} & \ct{3} & \ct{566} & \ct{190} & \ct{13.36} & \ct{4.367} \\
    \bottomrule
    \end{tabular}
    \label{tab:bench_char}
\end{table}

\textbf{FuCE}: We run \ac{FuCE} on S3C benchmarks using randomly generated input testcases. We set $time_{threshold}=5$s and $time_{budget}=1800$s for our experiments as per \ac{FuCE} (\autoref{algo:FuCE_1}). These are user defined configurable parameters.

We evaluate \ac{FuCE} with our baseline test generation techniques on two dimension: 1) Trojan detection capability 2) Branch coverage achievable during a pre-defined time limit. The first objective assumes availability of input-output response pairs from golden model to check Trojans functionally corrupted the design. However, the second objective aims to study the efficacy of test generation framework to achieve complete branch coverage on the design. Test-cases covering all conditional statements in the design enhance defenders' confidence of capturing any anomalous behaviour in  the absence of golden model. For both experimental settings, we present case-studies demonstrating the effectiveness of \ac{FuCE} on S3C benchmarks. For apples-to-apples comparison with prior state-of-art approaches \cite{symbolicSystemC,fuzzSystemC}, we compare the results of Trojan detection as reported in their published works.

\section{Empirical analysis}
\label{analysis}
\subsection{Trojan detection}

We first analyze trojan detection capability of \ac{FuCE} on S3C benchmarks. Since \ac{FuCE} invokes fuzzing and concolic engine interchangeably, we term each fuzzing phase as $fuzz_{n}$ and concolic execution phase as $conc_{n}$; where $n$ denotes phase ID in \ac{FuCE} execution. For example, a trojan detected in phase $fuzz_{3}$ implies that the framework has gone through test generation phases $fuzz_{1}$-$conc_{1}$-$fuzz_{2}$-$conc_{2}$-$fuzz_{3}$ before detecting the Trojan.

We evaluated Trojan detection capability of \ac{FuCE} on S3C benchmarks since it is the state-of-art trojan infected high-level designs. We reported the results in \autoref{tab:trojan_detection} and \ref{tab:trojanDetectionFuCE}. \autoref{tab:trojan_detection} shows testcase generated and runtime of each execution phase of \ac{FuCE}. We compare branch coverage obtained by \ac{FuCE} with \ac{AFL} and \ac{S2E} in \autoref{tab:trojan_detection}. In ~\autoref{tab:trojanDetectionFuCE}, we report the total time taken, the number of testcases generated for Trojan detection and memory usage of \ac{FuCE} and compare these with our baseline techniques (\ac{AFL} and \ac{S2E}) as well as with the state-of-art test-based Trojan detection approaches~\cite{fuzzSystemC,symbolicSystemC}. We set a timeout of two hours for trojan detection using each technique.

\begin{table}[t]
    \centering
    \small
    \caption{Trojan Detection by {\it FuCE}. Trojan types: \rt{CWOM}, \ct{SWM} and \mat{SWOM}. Detection: Yes (\cmark), No (\xmark)}
    \begin{tabular}{cccccccc}
        \toprule
        \multirow{2}{*}{\textbf{Benchmarks}} & \multicolumn{2}{c}{\textbf{Testcases}} & \multicolumn{2}{c}{\textbf{Time(in s)}} & \multicolumn{3}{c}{\textbf{Branch cov. (\%)}}\\
        \cmidrule(lr){2-3}\cmidrule(lr){4-5}\cmidrule(lr){6-8}
        & {$fuzz_1$} & {$conc_1$} & {$fuzz_1$} & {$conc_1$} & AFL & S2E & FuCE\\ \midrule
        \multirow{2}{*}{ADPCM} & \ct{3} & - & \ct{38} & - & \ct{88.1(\cmark)} & \ct{88.9(\cmark)} &\ct{88.1(\cmark)} \\
         & \mat{6} & - & \mat{15} & - & \mat{85.7(\cmark)} & \mat{86.1(\cmark)} &\mat{85.7(\cmark)} \\ \cmidrule(lr){1-8}
        AES & \rt{4} & \rt{1} & \rt{43} & \rt{4} & \rt{93.8(\cmark)} & \rt{81.5(\xmark)} &\rt{94.9(\cmark)} \\ \cmidrule(lr){1-8}
        \multirow{2}{*}{Bubble\_sort} & \rt{4} & \rt{8} & \rt{19} & \rt{192} & \rt{95.5(\xmark)} & \rt{95.5(\xmark)} &\rt{100(\cmark)} \\
         & \ct{3} & \ct{3} & \ct{19} & \ct{124} & \ct{95.5(\xmark)} & \ct{95.5(\xmark)} &\ct{100(\cmark)}  \\  \cmidrule(lr){1-8}
        Filter\_FIR & \rt{4} & \rt{2} & \rt{11} & \rt{13} & \rt{93.8(\cmark)} & \rt{93.8(\cmark)} & \rt{93.8(\cmark)} \\  \cmidrule(lr){1-8}
        \multirow{3}{*}{Interpolation} & \rt{63} & \rt{4} & \rt{11} & \rt{38} & \rt{45.1(\xmark)} & \rt{46(\xmark)} &\rt{76.1(\cmark)} \\ 
        & \ct{3} & - & \ct{4} & - & \ct{57.5(\cmark)} & \ct{56.1(\cmark)} &\ct{57.5(\cmark)} \\
        & \mat{3} & - & \mat{4} & - & \mat{57.5(\cmark)} & \mat{56.1(\cmark)} &\mat{57.5(\cmark)} \\\cmidrule(lr){1-8}
        Decimation & \ct{4} & - & \ct{24} & - & \ct{66.7(\cmark)} & \ct{66.7(\cmark)} &\ct{69.1(\cmark)} \\\cmidrule(lr){1-8}
        \multirow{2}{*}{Kasumi} & \rt{23} & - & \rt{5} & - & \rt{87.5(\cmark)} & \rt{84.3(\cmark)} & \rt{87.5(\cmark)} \\
        & \ct{22} & - & \ct{5} & - & \ct{87.5(\cmark)} & \ct{84.3(\cmark)} &\ct{87.5(\cmark)} \\\cmidrule(lr){1-8}
        UART-1 & \ct{6} & \ct{3} & \ct{34} & \ct{234} & \ct{85.7(\cmark)} & \ct{81.2(\xmark)} &\ct{88.5(\cmark)} \\
        UART-2 & \ct{2} & \ct{2} & \ct{32} & \ct{242} & \ct{79.4(\cmark)} & \ct{88.3(\cmark)} &\ct{88.3(\cmark)} \\
        \bottomrule
    \end{tabular}
    \label{tab:trojan_detection}
\end{table}

\noindent{\textbf{1) Coverage obtained till Trojan detection:}} We report the branch coverage results of \ac{FuCE} in \autoref{tab:trojan_detection}. We present the number of  test-cases and the time taken by \ac{FuCE} in each execution phase before Trojan detection. One important  point to note: \ac{FuCE} could detect all the Trojans in S3C benchmarks within $conc_1$ execution phase. The designs for which \ac{AFL} standalone can detect the trojan in $fuzz_1$ phase only without violating the threshold time, $t_{threshold}$, we do not invoke $conc_1$ phase. We present the branch coverage obtained by \ac{FuCE} and baseline techniques until Trojan detection (or, reaching pre-defined timeout limit).\\

\noindent{\textbf{Comparison with baselines:}}  We compare the branch coverage obtained by \ac{FuCE} and baseline techniques in \autoref{tab:trojan_detection} and observe that  \ac{FuCE} outperforms baseline techniques in terms of Trojan detection capability and coverage achieved till Trojan detection. For \textit{ADPCM}, \ac{FuCE} could detect the trojan in $fuzz_1$ phase itself. It is observed that \ac{AFL} and \ac{FuCE} perform similarly in trojan detection for \textit{ADPCM}. \ac{S2E} on the other hand detects trojan with little better coverage but takes longer time. Timing comparison is shown in~\autoref{tab:trojanDetectionFuCE}. For \textit{AES}, \ac{FuCE} goes through the phases $fuzz_1$ and $conc_1$ to detect trojan with better coverage than \ac{AFL} and \ac{S2E}. \ac{AFL} could not detect the trojan without violating the threshold time allocated for it to check for test inputs generated that hit new branch in the design. So, \ac{FuCE} shifts from $fuzz_1$ phase to $conc_1$ phase thus detecting trojan in less time than \ac{AFL} alone. For benchmarks \textit{AES}, \textit{Bubble\_sort}, \textit{Filter\_FIR}, \textit{Interpolation} and \textit{UART}, \ac{FuCE} invokes $conc_1$ indicating fuzz engine was stuck at $t_{threshold}$. Similarly, S2E could not detect all the Trojans in user-defined time limit indicating concolic execution is a slow process because of timing overhead from expensive SAT calls of concolic engine. An important advantage of using fuzzing alongside concolic execution is automatic identification of variable states which are responsible for complex checking operations. This effectively reduces the  burden on concolic engine to categorise the variables between symbolic and concrete before invoking it. In the next subsection, we outline case-studies on S3C benchmarks describing how \ac{FuCE} alleviates the challenges coming from standalone techniques to explore design state-space without hitting scalability bottleneck.\\

\noindent{\textbf{2) Analyzing timing improvement:}}
In our work, we compare wall-time for trojan detection of \ac{FuCE} with baseline and state-of-art techniques. The reason is that the fuzz engine takes less cpu-time for a given wall-time as it is an IO intensive process, whereas a concolic engine takes more cpu-time for a given wall-time as it forks multiple threads for test generation. For a fair comparison across a range of techniques, we choose wall-time as a metric to compare with previous works.  The wall-time taken for trojan detection is presented in \autoref{tab:trojanDetectionFuCE} (Column 3). {\em TO} indicates trojan is not detected within the wall-time limit of two hours.\\

\noindent{\textbf{Comparison with baselines:}} From \autoref{tab:trojanDetectionFuCE}, we conclude that \ac{FuCE} takes less time than vanilla \ac{AFL} for half of the benchmark designs considered and outperforms \ac{S2E} on all the designs except on \textit{Decimation}. \ac{FuCE} avoids expensive path exploration by concolic execution using fuzzer generated seeds leading to faster and scalable Trojan detection.\\

\begin{table}[t]
\centering
\caption{Comparing Trojan detection using \emph{FuCE} with prior works~\cite{fuzzSystemC,symbolicSystemC}. Trojan: \rt{CWOM}, \ct{SWM} and \mat{SWOM}.}
\resizebox{\columnwidth}{!}{
\begin{tabular}{lcccccccccccccc}
\toprule
\multicolumn{1}{c}{\multirow{3}{*}{\bf{Benchmarks}}} & \multicolumn{5}{c}{\bf{Test-cases generated}} & \multicolumn{5}{c}{\bf{Wall-time taken (s)}} & \multicolumn{4}{c}{\bf{Memory footprint(MB)}} \\ 
\cmidrule(lr){2-6} \cmidrule(lr){7-11}\cmidrule(lr){12-14}
& \begin{tabular}[c]{@{}l@{}}\bf{AFL}\end{tabular} & \begin{tabular}[c]{@{}l@{}}\bf{AFL-}\\\textbf{SHT}\end{tabular} & \begin{tabular}[c]{@{}l@{}}\\\bf{S2E}\end{tabular} & \begin{tabular}[c]{@{}l@{}}\bf{SCT-}\\ \bf{HTD}\end{tabular} & \begin{tabular}[c]{@{}l@{}}\\ \bf{FuCE}\end{tabular} & \bf{AFL} & \begin{tabular}[c]{@{}l@{}}\textbf{AFL-}\\ \textbf{SHT}\end{tabular} & \bf{S2E} & \begin{tabular}[c]{@{}l@{}}\bf{SCT-}\\ \bf{HTD}\end{tabular} & \bf{FuCE} & \begin{tabular}[c]{@{}l@{}}\bf{S2E}\\
\end{tabular} & \begin{tabular}[c]{@{}l@{}}\bf{SCT-}\\ \bf{HTD}\end{tabular} & \begin{tabular}[c]{@{}l@{}}\bf{FuCE}\\
\end{tabular} \\
\cline{1-15}

        \multirow{2}{*}{ADPCM} & \ct{3} & \ct{423} & \ct{14} & \ct{27} & \ct{3} & \ct{38} & \ct{1.17} & \ct{55} & \ct{157} & \ct{38} & \ct{3029} & \ct{3546} & \ct{51.44} \\
         & \mat{6} & \mat{414} & \mat{23} & \mat{7} & \mat{6} & \mat{15} & \mat{1.67} & \mat{49} & \mat{31} & \mat{15} & \mat{3433} & \mat{1341} & \mat{51.53} \\ \cmidrule(lr){1-15}
        AES & \rt{7} & \rt{22} & \rt{2} & \rt{11} & \rt{5} & \rt{337} & \rt{0.04} & \rt{TO} & \rt{23} & \rt{47} & \rt{9879} & \rt{1386} & \rt{1051}  \\ \cmidrule(lr){1-15}
        \multirow{2}{*}{Bubble\_sort} & \rt{30} & \rt{39} & \rt{160} & \rt{2} & \rt{12}  & \rt{TO} & \rt{4.82} & \rt{TO} & \rt{8} & \rt{211} & \rt{4761} & \rt{1074} & \rt{2698}  \\ 
        & \ct{12} & \ct{108} & \ct{32} & \ct{4} & \ct{6} & \ct{TO} & \ct{337.36} & \ct{TO} & \ct{10} & \ct{143} & \ct{4759}  & \ct{1106} & \ct{2618} \\ \cmidrule(lr){1-15}
        Filter\_FIR & \rt{11} & \rt{41} & \rt{5} & \rt{26} & \rt{6} & \rt{1184} & \rt{0.07} & \rt{41} & \rt{13} & \rt{24} & \rt{640} & \rt{1071} & \rt{480}  \\ \cmidrule(lr){1-15}
        \multirow{3}{*}{Interpolation} & \rt{63} & \rt{2325402} & \rt{72} & - & \rt{67} & \rt{TO} & \rt{TO} & \rt{TO} & - & \rt{49}   & \rt{16244} & - & \rt{3081}\\
        & \ct{3} & \ct{47} & \ct{8} & - & \ct{3} & \ct{4} & \ct{0.16} & \ct{130} & - & \ct{4} & \ct{8790} & - & \ct{56.20} \\
        & \mat{3} & \mat{47}  & \mat{2} & - & \mat{3} & \mat{4} & \mat{0.16} & \mat{14} & - & \mat{4} & \mat{3326} & - & \mat{56.21}\\
        \cmidrule(lr){1-15}
        Decimation & \ct{4} & - & \ct{2} & - & \ct{4} & \ct{22} & - & \ct{12} & - & \ct{24} & \ct{723} & -  & \ct{55.57}\\
        \cmidrule(lr){1-15}
        \multirow{2}{*}{Kasumi} & \rt{23} & \rt{316} & \rt{76} & - & \rt{23} & \rt{5} & \rt{1.32} & \rt{245} & - & \rt{5} & \rt{2908} & - & \rt{53.29}\\
        & \ct{22} & \ct{414} & \ct{49} & - & \ct{22} & \ct{5} & \ct{1.32} & \ct{83} & - & \ct{5} & \ct{2976} & - & \ct{53.28}\\ \cmidrule(lr){1-15}
        UART-1 & \ct{7} & \ct{51} & \ct{15} & \ct{3} & \ct{9} & \ct{311} & \ct{0.18} & \ct{TO} & \ct{9} & \ct{268} & \ct{5929} & \ct{1071} & \ct{3164}\\
        UART-2 & \ct{4} & - & \ct{2} & \ct{3} & \ct{4} & \ct{298} & - & \ct{730} & \ct{9} & \ct{274} & \ct{2651} & \ct{1070} & \ct{2618}\\
\bottomrule
\end{tabular}
}
\label{tab:trojanDetectionFuCE}
\end{table}

\noindent{\textbf{Comparison with state-of-art approaches:}} We reached out to the authors of \cite{fuzzSystemC,symbolicSystemC} to obtain the test generation frameworks for independent evaluation and apples-to-apples comparison with \ac{FuCE}. However, the actual implementations were unavailable. Thus, we compared the timing  as reported in their papers and used similar computing platform for our experiments. We found \ac{FuCE} took longer time than \ac{AFL-SHT} except for two cases: \textit{interpolation-cwom} and \textit{bubble\_sort-swm}; where \ac{FuCE} detects the trojan in 49s and 243s respectively. In these two cases, \ac{AFL-SHT} either took longer time or timed out. This exhibits fundamental limitation of fuzzing even though that fuzz engine is heavily customized based on benchmark characteristics. \ac{FuCE} on the other hand can easily identify the state of fuzz engine getting stuck and invoke concolic engine to penetrate into deeper program state. In a subsequent section, we will elaborately describe Trojan behaviour in S3C benchmarks and \ac{FuCE}'s ability to detect these. \ac{FuCE} detected Trojans quicker than \ac{SCT-HTD} for \textit{ADPCM}, \textit{Interpolation}, \textit{Decimation} and \textit{Kasumi} because \textit{SCT-HTD} took longer time to solve computationally intensive operations to generate the testcases.\\

\noindent{\textbf{3) Analyzing test-case quality:}} As listed in \autoref{tab:trojanDetectionFuCE} (Column 2), \ac{FuCE} leverages \ac{S2E} with fuzz generated test-cases to accelerate the coverage over a defined period of time. The fuzzer generated input seeds guide the symbolic engine to construct the execution tree along the execution path triggered by existing test-cases and generate new test cases reaching unexplored conditional statements. For a fair comparison with \ac{FuCE}, we report the number of test-cases preserved by each technique until it reaches user-defined timeout (or, till Trojan detection).\\

\noindent{\textbf{Comparison with baselines:}} Compared to \ac{AFL} and \ac{S2E}, the number of test-cases generated until Trojan detection are comparable for all benchmarks. A closer analysis reveals that the number of testcases generated by \ac{FuCE} is same as \ac{AFL} for the cases where \ac{AFL} as standalone was sufficient in detecting the Trojans without getting stuck for $time_{threshold}$. But, for cases where \ac{AFL} crossed the $time_{threshold}$ to generate a new testcase that improves coverage, \ac{FuCE} invoked concolic engine for generating qualitative test-cases quickly. Finally, \ac{FuCE} was able to detect Trojan using fewer test-cases than both \ac{AFL} and \ac{S2E}. This indicates that \ac{FuCE}  uses the time-budget judiciously for creating effective test-cases that explore deeper program segments when fuzzing is stuck.\\

\noindent{\textbf{Comparison with state-of-art approaches}}: We compared the number of test-cases generated via state-of-art approaches with \ac{FuCE}.  \ac{FuCE} outperforms \ac{SCT-HTD} for designs like \textit{ADPCM}, \textit{AES} and \textit{Filter-FIR} by generating fewer test-cases. Although \ac{SCT-HTD} is better than \ac{FuCE} in terms of generating effective test-cases for designs like \textit{Bubble-sort} and \textit{UART}, we will explore in subsequent sections that this heavily depends on the exploration strategy selected by concolic engine to explore the search space. Due to unavailability of branch coverage by \ac{SCT-HTD} till Trojan detection, it is difficult to conclude whether \ac{SCT-HTD} generated fewer test-cases with better coverage, or explored Trojan location quickly and terminated before achieving substantial coverage on the rest of the design.\\

\noindent{\textbf{4) Analyzing memory footprint:}} The last column of \autoref{tab:trojanDetectionFuCE} reports memory usage denoting maximum memory footprint . We observe that trojans are detected with reasonable memory usage by \ac{FuCE}. We compare it with concolic execution engine \ac{S2E} and \ac{SCT-HTD} reported in ~\cite{symbolicSystemC}. \emph{AFL} is an input-output~(IO) bound process whereas \ac{S2E} is memory intensive program allocating huge memory for an application to run on a virtual machine. For \ac{AFL} and \ac{AFL-SHT}, we used default configuration of 50MB memory which was sufficient for all the S3C benchmarks. From the results, one can interpret that fuzzing combined with concolic execution has 50\% less memory footprint(average) compared to standalone concolic execution.


\input{images/fig_branchCov}

\subsection{Coverage Improvement}

For analyzing the effectiveness of \ac{FuCE} framework, we measure the branch coverage obtained by running the baseline techniques and \ac{FuCE} with a time-limit of two hours.  \autoref{fig:coverageOfFuCE} and \autoref{tab:cov_improve} study the detailed coverage analysis over the entire time period for \ac{FuCE} and baseline techniques. 

From the coverage data (~\autoref{fig:coverageOfFuCE}), we can categorize S3C benchmarks into two types from testing perspective: simple and complex. For simple designs having small number of nested conditional statements, \ac{FuCE} could achieve 100\% coverage within a short span of time with fewer test-cases. These are: \textit{Bubble-sort} and \textit{Kasumi}. Complex benchmarks have deeper levels of nested loop and conditional statements along with ternary operations. These are: \textit{ADPCM} and \textit{Interpolation}. 
\ac{FuCE} achieved 100\% branch coverage on all S3C benchmarks except \textit{UART}, \textit{AES}, \textit{Filter\_FIR} and \textit{Decimation}. We dig deeper into the coverage analysis for which \ac{FuCE} could not achieve 100\% coverage and discovered an interesting observation: presence of unreachable branch conditions in the original benchmark designs. We analytically found uncovered branch conditions from our experiments and verified that no test-case is possible for covering these code segments. We believe that these unreachable code segments will be optimized out during \ac{RTL} level synthesis and therefore is not a drawback of \ac{FuCE} test generation framework.

\begin{table*}
    \centering
    \caption{Coverage improvement by {\it FuCE}. Trojan types: \rt{CWOM}, \ct{SWM} and \mat{SWOM}.}
   \resizebox{0.85\columnwidth}{!}{
    \begin{tabular}{ccccccccc}
        \toprule
        \multirow{2}{*}{\textbf{Benchmarks}} &
        \multicolumn{3}{c}{\textbf{\#Testcases generated}} &
        \multicolumn{3}{c}{\textbf{Time(in s)}} &  \multirow{2}{*}{\textbf{Phases}} & \multirow{2}{*}{\textbf{Branch cov. (\%)}} \\
        \cmidrule(lr){2-4} \cmidrule(lr){5-7}
        & {$fuzz_1$} & {$conc_1$} & {$fuzz_2$} & {$fuzz_1$} & {$conc_1$} & {$fuzz_2$} & & \\ 
        \midrule
         \multirow{2}{*}{ADPCM} & \ct{3} & \ct{39} & \ct{3} & \ct{43} & \ct{1800} & \ct{1940} & \ct{$fuzz_1$-$conc_1$-$fuzz_2$} & \ct{100}\\
        & \mat{6} & \mat{82} & \mat{4} & \mat{20} & \mat{1800} & \mat{1640} & \mat{$fuzz_1$-$conc_1$-$fuzz_2$} & \mat{100} \\
        \cmidrule(lr){1-9}
        AES & \rt{4} & \rt{1} & - & \rt{43} & \rt{4} & - & \rt{$fuzz_1$-$conc_1$} & \rt{94.9} \\
        \cmidrule(lr){1-9}
        \multirow{2}{*}{Bubble\_sort} & \rt{4} & \rt{8} & - & \rt{19} & \rt{192} & - & \rt{$fuzz_1$-$conc_1$} & \rt{100} \\
        & \ct{3} & \ct{3} & - & \ct{19} & \ct{124} & - & \ct{$fuzz_1$-$conc_1$} & \ct{100} \\
        \cmidrule(lr){1-9}
        Filter\_FIR & \rt{4} & \rt{2} & - & \rt{13} & \rt{11} & - & \rt{$fuzz_1$-$conc_1$} & \rt{93.8}\\
        \cmidrule(lr){1-9}
        \multirow{3}{*}{Interpolation} & \rt{63} & \rt{460} & \rt{55} & \rt{11} & \rt{1800} & \rt{3880} & \rt{$fuzz_1$-$conc_1$-$fuzz_2$} & \rt{100}\\
        & \ct{3} & \ct{1427} & \ct{28} & \ct{9} & \ct{270} & \ct{2217} & \ct{$fuzz_1$-$conc_1$-$fuzz_2$} & \ct{100}\\
        & \mat{3} & \mat{663} & \mat{51} & \mat{9} & \mat{270} & \mat{3640} & \mat{$fuzz_1$-$conc_1$-$fuzz_2$} & \mat{100}\\
         \cmidrule(lr){1-9}
        Decimation & \ct{4} & \ct{12} & \ct{5} & \ct{29} & \ct{126} & \ct{1429} & \ct{$fuzz_1$-$conc_1$-$fuzz_2$} & \ct{96.8}\\
         \cmidrule(lr){1-9}
        \multirow{2}{*}{Kasumi} & \rt{23} & \rt{22} & - & \rt{10} & \rt{1221} & - & \rt{$fuzz_1$-$conc_1$} & \rt{100}\\
        & \ct{22} & \ct{12} & - & \ct{10} & \ct{1682} & - & \ct{$fuzz_1$-$conc_1$} & \ct{100}\\
        \cmidrule(lr){1-9}
        UART-1 & \ct{6} & \ct{3} & - & \ct{34} & \ct{234} & - & \ct{$fuzz_1$-$conc_1$} & \ct{88.5}\\
        UART-2 & \ct{2} & \ct{2} & - & \ct{32} & \ct{242} & - & \ct{$fuzz_1$-$conc_1$} & \ct{88.3}\\
        \bottomrule
    \end{tabular}
    }
    \label{tab:cov_improve}
\end{table*}

\subsection{Case Studies}

Here, we dive deep into \ac{FuCE}'s performance on four \emph{S3CBench} designs across two orthogonal directions: 1) Trojan  detection  capability and 2) achievable branch  coverage within a defined time  limit.\\

\noindent{\textbf{1) Case Study I: Trojan Detection}}

{\textit{Interpolation}} is a 4-stage interpolation filter. We consider \textit{CWOM} trojan variant for our case study. Only \ac{FuCE} could successfully detect the trojan amongst every other state-of-art technique (\autoref{tab:trojanDetectionFuCE}). We show \textit{CWOM} trojan variant of interpolation in~\autoref{label:interpcode}. The trojan triggers once the output of FIR filter's final stage ($SoP4$) matches with a specific ``magic" value. Line \ref{line:trigger} is Trigger logic and line \ref{line:payload} shows the payload circuit as output write operation. The trigger activates for a input resulting in the sum of product of the fourth filter($SoP4$) as $-.015985481441$. Trigger activation leads to execution of payload circuit  (line \ref{line:trigger}) and writes the output $odata-write = -0.26345$. Inputs not satisfying the Trigger condition behave functionally equivalent to Trojan-free design.

Fuzzing techniques like \ac{AFL} are unlikely to satisfy the conditional checks against specific values (line \ref{line:trigger} of~\autoref{label:interpcode}) in a short time-span. \ac{AFL} failed to detect the trojan in two hours time limit. \ac{S2E} executing with random seeds also failed to generate inputs satisfying the trojan trigger condition. However, \ac{FuCE} leverages the strength of both fuzzing and concolic execution. \ac{FuCE} passes the interesting inputs as identified by the fuzzer in phase $fuzz_1$ to S2E in phase $conc_1$. S2E traces each input generated by the fuzzer discovering unexplored program states by \ac{AFL} during phase $fuzz_1$. \ac{S2E} generates inputs satisfying complex branch conditions to explore the undiscovered states. Thus, \ac{FuCE} triggered the trojan payload using \ac{S2E} and successfully detected the Trojan. \\
\begin{lstlisting}[caption={Interpolation - Trojan Logic}, style=customc,label={label:interpcode}]
/*Trojan Triggers based on particular input combination */
#elif defined(CWOM) ||defined(CWOM_TRI) 
    if(SoP4 == -.015985481441)  ?\label{line:trigger}? //trojan trigger logic 
        odata_write = -0.26345; ?\label{line:payload}? //trojan payload circuit
    else
        //Trojan free execution for odata_write
#endif
\end{lstlisting}

\vspace{0.8cm}
{\textit{Advanced Encryption Standard}} (AES) is a symmetric block cipher algorithm. The plain text is 128 bits long and key can be 128, 192 or 256 bits. \emph{S3CBench} contain AES-128 bit design and the trojan type is \textit{CWOM}. This trojan leaks the secret key for a specific plain-text input corrupting the encryption generating incorrect cipher-text. \\
 AES-128 performs ten rounds of repetitive operation to generate cipher-text $CT_{10}$. The trojan implementation performs an additional 'n' rounds to generate cipher-text ($CT_{10+n}$). Initially, the attacker tampers the key $K_n$ to be used in the round $R_{10+n}$ for the round operations: \textit{SubBytes, ShiftRows, MixColumns, AddRoundKey}. Later, key $K_{10}$ can be recovered from the plain text $P$ and cipher-text $CT_{10+n}$. Using \ac{FuCE}, we compare the generated cipher-text with the expected cipher-text to detect the presence of a Trojan. 
 
\begin{lstlisting}[caption={AES - Trojan Logic}, style=customc,label={label:aescode}]
//Trojan Triggers for a particular input present in the plaintext message 	
#ifdef CWOM 
    if (data3[0] == -1460255950) // trojan trigger logic
        odata[(i * NB) + j] = (data1[i] $>>$ (j * 8));// trojan payload circuit
    else  //Trojan free execution for output data
        odata[(i * NB) + j] = 0;
#endif
\end{lstlisting}
\vspace{0.5cm}
 We examined the performance of \ac{FuCE} on \textit{AES}. \ac{FuCE} successfully detected the Trojan in \textit{AES-cwom} in phase $fuzz_1$-$conc_1$ with a $6.5$x timing speed-up compared to \ac{AFL}. From~\autoref{tab:trojan_detection}, we observe that \ac{FuCE} detected the trojan using the testcase generated by \ac{S2E} during phase $conc_1$. We supplied four test-cases generated by \ac{AFL} during $fuzz_1$ to \ac{S2E}. However, \ac{S2E} supplied with random test-cases was unable to detect the trojan (\autoref{tab:trojanDetectionFuCE}) and timed-out with a branch coverage of 81.5\%. We indicate the Trojan infected \textit{AES} in~\autoref{label:aescode} (lines 2 and 3). The Trojan trigger condition for \emph{AES} is a rare combination of input values. \\



\noindent{\textbf{2) Case Study II: Coverage Improvement}}

\begin{lstlisting}[caption={ADPCM - Encode and Decode Logic}, style=customc,label={label:adpcmcode}]
//Encode Logic
if (diff[15] == 1) {    //checking exactness of value
    width.diff_data = ((diff ^ 0xffff) + 1);
    neg_flag = true;
}
else {
    width.diff_data = diff;
    neg_flag = false;
}
//Decode Logic
 if (width.enc_data > 7) {
    width.enc_data = 7;
    dec_tmp = width.enc_data * divider;
    remainder1 = dec_tmp.range(1, 0);
    if (remainder1 >= 2) {  //Nested branch conditions
        width.pre_data += (dec_tmp >> 2) + 1;
    }
    else {
        width.pre_data += (dec_tmp >> 2);
    }
}   
\end{lstlisting}
\vspace{0.5cm}
{\textit{Adaptive Differential Pulse Code Modulation}} (ADPCM) converts analog information to binary data. \emph{ADPCM} converts 16 bits Pulse Code Modulation (PCM) samples into 4-bit samples. The trojan considered is \emph{SWM} type. The trojan gets triggered once the counter reaches a specific value corrupting the modulation process. Although \ac{FuCE} detects the trojan successfully at phase $fuzz_1$ with a coverage of 88.1\% but does not attain 100\% coverage at this phase. From the \emph{LCOV} report it was observed that \ac{FuCE} was unable to reach certain portions of code with nested conditional branch statements as given in~\autoref{label:adpcmcode}. So \ac{FuCE} goes to phase $conc_1$ for further exploration with the input seeds generated by $fuzz_1$. At $conc_1$, \ac{S2E} traces the program following the same path taken by the fuzzer. When \ac{S2E} arrives at the conditional check at line 2 (~\autoref{label:adpcmcode}) of the encode logic, it realizes that the path was not covered by the fuzzer. So, it produces input satisfying the condition which drives the execution to this new state transition. After the phase $conc_1$, the coverage analyzer for \ac{FuCE} framework found that the coverage improved to 89.5\% but the target coverage of 100\% was yet to be reached. Since concolic execution is a slow process, \ac{S2E} fails to generate test inputs within its time bound, that satisfy the nested conditional statements at line 11 of the decode logic(~\autoref{label:adpcmcode}). So \ac{FuCE} goes to the next phase $fuzz_2$ to look for the undiscovered paths. \ac{AFL} starts its execution with the input seeds generated by \ac{S2E} at phase $conc_1$ which guides the fuzzer to quickly penetrate in the nested branch conditions and generate test cases that give 100\% branch coverage for \ac{FuCE}.



Thus \emph{FuCE} could achieve 100\% branch coverage following the phases $fuzz_1$-$conc_1$-$fuzz_2$ in the defined time budget of two hours. From the plot of \emph{ADPCM-SWM} in \autoref{fig:coverageOfFuCE} (a), we can see that \ac{FuCE} successfully achieves coverage of 100\% in 3800 seconds. AFL on the other hand achieves coverage of 96.4\% in 6115 seconds and could not improve further in the test time of two hours. S2E reaches a maximum coverage of 88.9\% at 310 seconds while running for two hours. The breakdown of result for code coverage of \emph{ADPCM-SWM} by \emph{FuCE} could be found in ~\autoref{tab:cov_improve}.\\

{\textit{Decimation Filter}} is a 5-stage filter with five FIR filters cascaded together. The type of trojan inserted here is \textit{SWM}, which is triggered for a particular value of count in the design. The trojan is inserted here in the final stage, i.e., the fifth stage. Results on evaluation of this design of \emph{S3CBench} have not been reported by any of the state-of-art prior works \ac{AFL-SHT} or \ac{SCT-HTD}. We have evaluated this benchmark successfully with \ac{FuCE}, \ac{AFL} and \ac{S2E}. In our experiments, all the techniques could successfully detect the trojan in the circuit but \ac{FuCE} outperformed others significantly with respect to achievable branch coverage. From the plot of \emph{decimation-SWM}, in~\autoref{fig:coverageOfFuCE} (f), it is observed that the seeds generated by \ac{FuCE} could attain a coverage 96.8\% running for 3289 seconds.  Whereas for \ac{AFL} the maximum coverage of 69.1\% is attained with seeds generated in the time interval of (30 - 60) seconds and \ac{S2E} reached the coverage of 73\% in the time interval (100 - 300) seconds beyond which coverage did not increase even after running for two hours of time limit. \ac{FuCE} could cover almost all the portions of code using its interlaced execution of phases $fuzz_1$-$conc_1$-$fuzz_2$ that \ac{AFL} and \ac{S2E} failed to cover individually.

\label{label:exResults}

%% file: images/fig_branchCov.tex
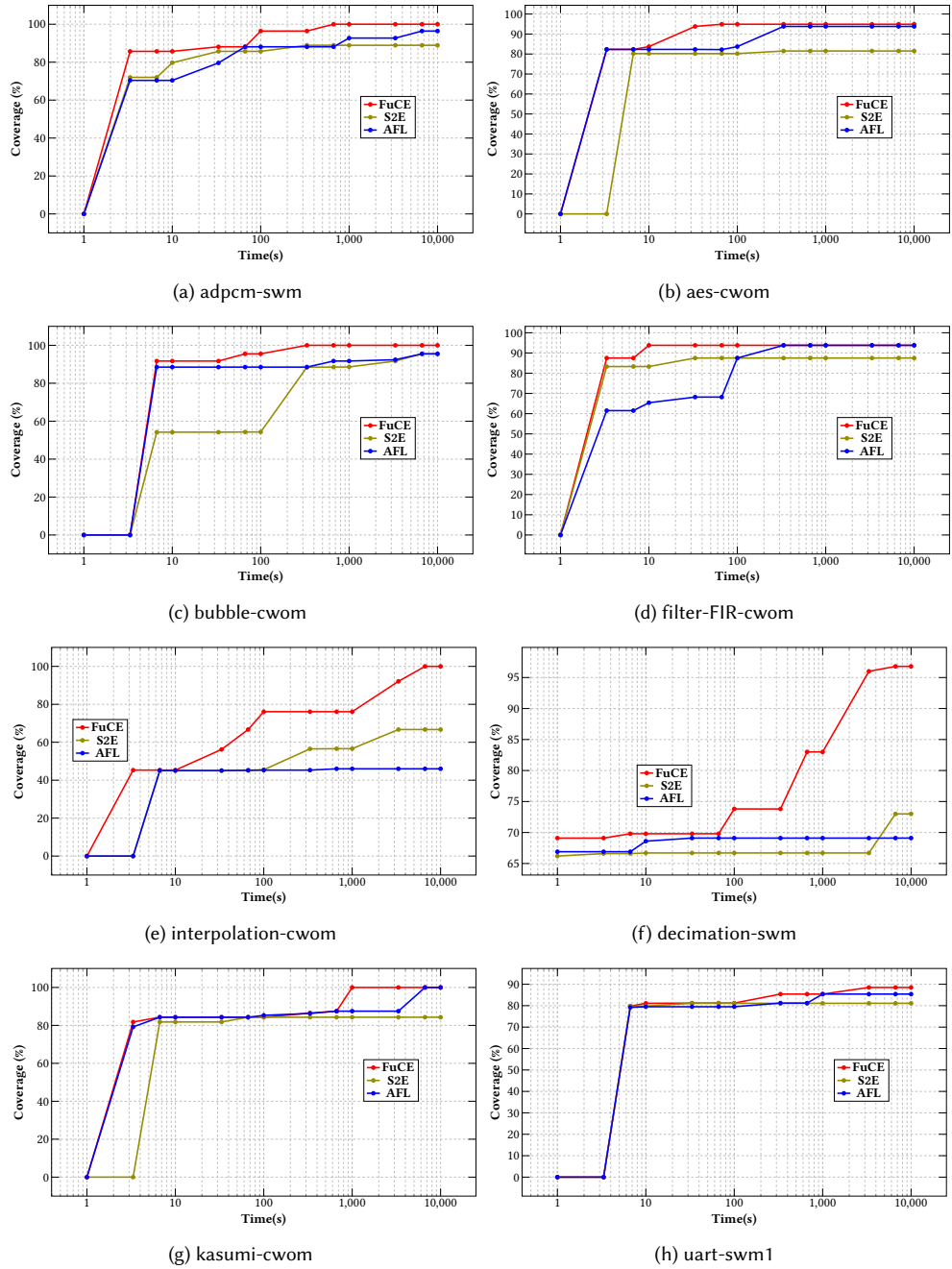
\begin{figure*}[h!]
\centering
    \subfloat[adpcm-swm]{
		\begin{tikzpicture}[scale=0.3] 
		\begin{axis}
			[    xmode=log,
            log ticks with fixed point,
			height=0.8\columnwidth,
			width=1.4\columnwidth,
			xlabel= Epsilon ($\epsilon$), 
			ylabel= Accuracy,
			every major tick/.append style={very thick, major tick length=10pt, black},
			axis line style = very thick,
			                    tick label style={font=\Huge},
			                                        label style={font=\Huge},
			                    xlabel=\textbf{Time(s)},ylabel=\textbf{Coverage (\%)},
			grid style = dashed,
			grid=both,
			legend style=
			{at={(0.8,0.6)}, 
				anchor=south west, 
				anchor= north , 
			} ,
			]
			\addplot[color=red,solid,mark=*,line width=2] plot coordinates {
				(1,     0)
				(3.33,  85.7)
				(6.67,  85.7)
				(10,    85.7)
				(33.3,  88.1)
				(66.6,  88.1)
				(100,   96.4)
                (333.33, 96.4)
                (666.67, 100)
                (1000,   100)
                (3333.33, 100)
                (6666.67, 100)
                (10000,   100)
			};
			
			\addplot[color=olive, solid, mark=*, mark size = 2, line width=2] plot coordinates {
				(1,     0)
				(3.33,  72.0)
				(6.67,  72.0)
				(10,    79.7)
				(33.3,  85.7)
				(66.6,  85.7)
				(100,   85.7)
                (333.33, 88.9)
                (666.67, 88.9)
                (1000,   88.9)
                (3333.33, 88.9)
                (6666.67, 88.9)
                (10000,   88.9)
			};
			\addplot[color=blue, solid, mark=*, line width=2] plot coordinates {
				(1,     0)
				(3.33,  70.4)
				(6.67,  70.4)
				(10,    70.4)
				(33.3,  79.6)
				(66.6,  88.1)
				(100,   88.1)
                (333.33, 88.1)
                (666.67, 88.1)
                (1000,   92.7)
                (3333.33, 92.7)
                (6666.67, 96.4)
                (10000,   96.4)
			};
		\legend{\Huge{\bf{FuCE}}\\\Huge{\bf{S2E}}\\\Huge{\bf{AFL}}\\}
			\end{axis}
		\end{tikzpicture}
	}
	\subfloat[aes-cwom]{
		\begin{tikzpicture}[scale=0.3] 
		\begin{axis}
			[    xmode=log,
            log ticks with fixed point,
			height=0.8\columnwidth,
			width=1.4\columnwidth,
			xlabel= Epsilon ($\epsilon$), 
			ylabel= Accuracy,
			every major tick/.append style={very thick, major tick length=10pt, black},
			axis line style = very thick,
			                    tick label style={font=\Huge},
			                                        label style={font=\Huge},
			                    xlabel=\textbf{Time(s)},ylabel=\textbf{Coverage (\%)},
			grid style = dashed,
			grid=both,
			legend style=
			{at={(0.8,0.6)}, 
				anchor=south west, 
				anchor= north , 
			} ,
			]
			\addplot[color=red,solid,mark=*,line width=2] plot coordinates {
				(1,     0)
				(3.33,  82.3)
				(6.67,  82.3)
				(10,    83.7)
				(33.3,  93.8)
				(66.6,  94.9)
				(100,   94.9)
                (333.33, 94.9)
                (666.67, 94.9)
                (1000,   94.9)
                (3333.33, 94.9)
                (6666.67, 94.9)
                (10000,   94.9)
			};
			\addplot[color=olive, solid, mark=*, mark size = 2, line width=2] plot coordinates {
				(1,     0)
				(3.33,  0)
				(6.67,  80.2)
				(10,    80.2)
				(33.3,  80.2)
				(66.6,  80.2)
				(100,   80.2)
                (333.33, 81.5)
                (666.67, 81.5)
                (1000,   81.5)
                (3333.33, 81.5)
                (6666.67, 81.5)
                (10000,   81.5)
			};
			\addplot[color=blue, solid, mark=*, line width=2] plot coordinates {
				(1,     0)
				(3.33,  82.3)
				(6.67,  82.3)
				(10,    82.3)
				(33.3,  82.3)
				(66.6,  82.2)
				(100,   83.7)
                (333.33, 93.8)
                (666.67, 93.8)
                (1000,   93.8)
                (3333.33, 93.8)
                (6666.67, 93.8)
                (10000,   93.8)
			};
		\legend{\Huge{\bf{FuCE}}\\\Huge{\bf{S2E}}\\\Huge{\bf{AFL}}\\}
			\end{axis}
		\end{tikzpicture}
	} 
	
	\vspace{0.1in}
	
	 \subfloat[bubble-cwom]{
		\begin{tikzpicture}[scale=0.3] 
		\begin{axis}
			[    xmode=log,
    log ticks with fixed point,
			height=0.8\columnwidth,
			width=1.4\columnwidth,
			xlabel= Epsilon ($\epsilon$), 
			ylabel= Accuracy,
			every major tick/.append style={very thick, major tick length=10pt, black},
			axis line style = very thick,
			                    tick label style={font=\Huge},
			                                        label style={font=\Huge},
			                    xlabel=\textbf{Time(s)},ylabel=\textbf{Coverage (\%)},
			grid style = dashed,
			grid=both,
			legend style=
			{at={(0.8,0.6)}, 
				anchor=south west, 
				anchor= north , 
			} ,
			]
			\addplot[color=red,solid,mark=*,line width=2] plot coordinates {
				(1,     0)
				(3.33,  0)
				(6.67,  91.7)
				(10,    91.7)
				(33.3,  91.7)
				(66.6,  95.5)
				(100,   95.5)
                (333.33, 100)
                (666.67, 100)
                (1000,   100)
                (3333.33, 100)
                (6666.67, 100)
                (10000,   100)
			};
			\addplot[color=olive, solid, mark=*, mark size = 2, line width=2] plot coordinates {
				(1,     0)
				(3.33,  0)
				(6.67,  54.2)
				(10,    54.2)
				(33.3,  54.2)
				(66.6,  54.3)
				(100,   54.3)
                (333.33, 88.5)
                (666.67, 88.5)
                (1000,   88.6)
                (3333.33, 91.7)
                (6666.67, 95.5)
                (10000,   95.5)
			};
			\addplot[color=blue, solid, mark=*, line width=2] plot coordinates {
				(1,     0)
				(3.33,  0)
				(6.67,  88.5)
				(10,    88.5)
				(33.3,  88.5)
				(66.6,  88.5)
				(100,   88.5)
                (333.33, 88.5)
                (666.67, 91.7)
                (1000,   91.7)
                (3333.33, 92.4)
                (6666.67, 95.5)
                (10000,   95.5)
			};
		\legend{\Huge{\bf{FuCE}}\\\Huge{\bf{S2E}}\\\Huge{\bf{AFL}}\\}
			\end{axis}
		\end{tikzpicture}
	} 
 	\subfloat[filter-FIR-cwom]{
		\begin{tikzpicture}[scale=0.3] 
		\begin{axis}
			[    xmode=log,
    log ticks with fixed point,
			height=0.8\columnwidth,
			width=1.4\columnwidth,
			xlabel= Epsilon ($\epsilon$), 
			ylabel= Accuracy,
			every major tick/.append style={very thick, major tick length=10pt, black},
			axis line style = very thick,
			                    tick label style={font=\Huge},
			                                        label style={font=\Huge},
			                    xlabel=\textbf{Time(s)},ylabel=\textbf{Coverage (\%)},
			grid style = dashed,
			grid=both,
			legend style=
			{at={(0.8,0.6)}, 
				anchor=south west, 
				anchor= north , 
			} ,
			]
			\addplot[color=red,solid,mark=*,line width=2] plot coordinates {
				(1,     0)
				(3.33,  87.5)
				(6.67,  87.5)
				(10,    93.8)
				(33.3,  93.8)
				(66.6,  93.8)
				(100,   93.8)
                (333.33, 93.8)
                (666.67, 93.8)
                (1000,   93.8)
                (3333.33, 93.8)
                (6666.67, 93.8)
                (10000,   93.8)
			};
			\addplot[color=olive, solid, mark=*, mark size = 2, line width=2] plot coordinates {
				(1,     0)
				(3.33,  83.3)
				(6.67,  83.3)
				(10,    83.3)
				(33.3,  87.5)
				(66.6,  87.5)
				(100,   87.5)
                (333.33, 87.5)
                (666.67, 87.5)
                (1000,   87.5)
                (3333.33, 87.5)
                (6666.67,87.5)
                (10000,   87.5)
			};
			\addplot[color=blue, solid, mark=*, line width=2] plot coordinates {
				(1,     0)
				(3.33,  61.5)
				(6.67,  61.5)
				(10,    65.4)
				(33.3,  68.2)
				(66.6,  68.2)
				(100,   87.5)
                (333.33, 93.8)
                (666.67, 93.8)
                (1000,   93.8)
                (3333.33, 93.8)
                (6666.67, 93.8)
                (10000,   93.8)
			};
		\legend{\Huge{\bf{FuCE}}\\\Huge{\bf{S2E}}\\\Huge{\bf{AFL}}\\}
			\end{axis}
		\end{tikzpicture}
	}
	\vspace{0.1in}
	\subfloat[interpolation-cwom]{
		\begin{tikzpicture}[scale=0.3] 
		\begin{axis}
			[    xmode=log,
    log ticks with fixed point,
			height=0.8\columnwidth,
			width=1.4\columnwidth,
			xlabel= Epsilon ($\epsilon$), 
			ylabel= Accuracy,
			every major tick/.append style={very thick, major tick length=10pt, black},
			axis line style = very thick,
			                    tick label style={font=\Huge},
			                                        label style={font=\Huge},
			                    xlabel=\textbf{Time(s)},ylabel=\textbf{Coverage (\%)},
			grid style = dashed,
			grid=both,
			legend style=
			{at={(0.05,0.5)}, 
				anchor=south west, 
				anchor= south west , 
			} ,
			]
			\addplot[color=red,solid,mark=*,line width=2] plot coordinates {
				(1,     0)
				(3.33,  45.3)
				(6.67,  45.3)
				(10,    45.3)
				(33.3,  56.2)
				(66.6,  66.7)
				(100,   76.1)
                (333.33, 76.1)
                (666.67, 76.1)
                (1000,   76.1)
                (3333.33, 92.1)
                (6666.67, 100.0)
                (10000,   100.0)
			};
			\addplot[color=olive, solid, mark=*, mark size = 2, line width=2] plot coordinates {
				(1,     0)
				(3.33,  0)
				(6.67,  45)
				(10,    45)
				(33.3,  45)
				(66.6,  45.2)
				(100,   45.5)
                (333.33, 56.5)
                (666.67, 56.6)
                (1000,   56.6)
                (3333.33, 66.7)
                (6666.67, 66.7)
                (10000,   66.7)
			};
			\addplot[color=blue, solid, mark=*, line width=2] plot coordinates {
				(1,     0)
				(3.33,  0)
				(6.67,  45.1)
				(10,    45.1)
				(33.3,  45.1)
				(66.6,  45.2)
				(100,   45.3)
                (333.33, 45.3)
                (666.67, 46)
                (1000,   46)
                (3333.33, 46)
                (6666.67, 46)
                (10000,   46)
			};
		\legend{\Huge{\bf{FuCE}}\\\Huge{\bf{S2E}}\\\Huge{\bf{AFL}}\\}
			\end{axis}
		\end{tikzpicture}
	}
\subfloat[decimation-swm]{
		\begin{tikzpicture}[scale=0.3] 
		\begin{axis}
			[    xmode=log,
    log ticks with fixed point,
			height=0.8\columnwidth,
			width=1.4\columnwidth,
			xlabel= Epsilon ($\epsilon$), 
			ylabel= Accuracy,
			every major tick/.append style={very thick, major tick length=10pt, black},
			axis line style = very thick,
			                    tick label style={font=\Huge},
			                                        label style={font=\Huge},
			                    xlabel=\textbf{Time(s)},ylabel=\textbf{Coverage (\%)},
			grid style = dashed,
			grid=both,
			legend style=
			{at={(0.4,0.3)}, 
				anchor=south west, 
				anchor= south east, 
			} ,
			]
			\addplot[color=red,solid,mark=*,line width=2] plot coordinates {
				(1,     69.1)
				(3.33,  69.1)
				(6.67,  69.8)
				(10,    69.8)
				(33.3,  69.8)
				(66.6,  69.8)
				(100,   73.8)
                (333.33, 73.8)
                (666.67, 83.0)
                (1000,   83.0)
                (3333.33, 96.0)
                (6666.67, 96.8)
                (10000,   96.8)
			};
			\addplot[color=olive, solid, mark=*, mark size = 2, line width=2] plot coordinates {
				(1,     66.2)
				(3.33,  66.6)
				(6.67,  66.6)
				(10,    66.7)
				(33.3,  66.7)
				(66.6,  66.7)
				(100,   66.7)
                (333.33, 66.7)
                (666.67, 66.7)
                (1000,   66.7)
                (3333.33, 66.7)
                (6666.67, 73.0)
                (10000,   73.0)
			};
			\addplot[color=blue, solid, mark=*, line width=2] plot coordinates {
				(1,     66.9)
				(3.33,  66.9)
				(6.67,  66.9)
				(10,    68.6)
				(33.3,  69.1)
				(66.6,  69.1)
				(100,   69.1)
                (333.33, 69.1)
                (666.67, 69.1)
                (1000,   69.1)
                (3333.33, 69.1)
                (6666.67, 69.1)
                (10000,   69.1)
			};
		\legend{\Huge{\bf{FuCE}}\\\Huge{\bf{S2E}}\\\Huge{\bf{AFL}}\\}
			\end{axis}
		\end{tikzpicture}
	}
	\vspace{0.1in}
	\subfloat[kasumi-cwom]{
		\begin{tikzpicture}[scale=0.3] 
		\begin{axis}
			[    xmode=log,
            log ticks with fixed point,
			height=0.8\columnwidth,
			width=1.4\columnwidth,
			xlabel= Epsilon ($\epsilon$), 
			ylabel= Accuracy,
			every major tick/.append style={very thick, major tick length=10pt, black},
			axis line style = very thick,
			                    tick label style={font=\Huge},
			                                        label style={font=\Huge},
			                    xlabel=\textbf{Time(s)},ylabel=\textbf{Coverage (\%)},
			grid style = dashed,
			grid=both,
			legend style=
			{at={(0.8,0.6)}, 
				anchor=south west, 
				anchor= north , 
			} ,
			]
			\addplot[color=red,solid,mark=*,line width=2] plot coordinates {
				(1,     0)
				(3.33,  81.8)
				(6.67,  84.3)
				(10,    84.3)
				(33.3,  84.3)
				(66.6,  84.3)
				(100,   84.3)
                (333.33, 86.5)
                (666.67, 87.5)
                (1000,   100.0)
                (3333.33, 100.0)
                (6666.67, 100.0)
                (10000,   100.0)
			};
			\addplot[color=olive, solid, mark=*, mark size = 2, line width=2] plot coordinates {
				(1,     0)
				(3.33,  0)
				(6.67,  81.8)
				(10,    81.8)
				(33.3,  81.8)
				(66.6,  84.3)
				(100,   84.3)
                (333.33, 84.3)
                (666.67, 84.3)
                (1000,   84.3)
                (3333.33, 84.3)
                (6666.67, 84.3)
                (10000,   84.3)
			};
			\addplot[color=blue, solid, mark=*, line width=2] plot coordinates {
				(1,     0)
				(3.33,  79.2)
				(6.67,  84.3)
				(10,    84.3)
				(33.3,  84.3)
				(66.6,  84.3)
				(100,   85.3)
                (333.33, 86.3)
                (666.67, 87.5)
                (1000,   87.5)
                (3333.33, 87.5)
                (6666.67, 100.0)
                (10000,   100.0)
			};
		\legend{\Huge{\bf{FuCE}}\\\Huge{\bf{S2E}}\\\Huge{\bf{AFL}}\\}
			\end{axis}
		\end{tikzpicture}
	}
    \subfloat[uart-swm1]{
		\begin{tikzpicture}[scale=0.3] 
		\begin{axis}
			[    xmode=log,
            log ticks with fixed point,
			height=0.8\columnwidth,
			width=1.4\columnwidth,
			xlabel= Epsilon ($\epsilon$), 
			ylabel= Accuracy,
			every major tick/.append style={very thick, major tick length=10pt, black},
			axis line style = very thick,
			                    tick label style={font=\Huge},
			                                        label style={font=\Huge},
			                    xlabel=\textbf{Time(s)},ylabel=\textbf{Coverage (\%)},
			grid style = dashed,
			grid=both,
			legend style=
			{at={(0.8,0.6)}, 
				anchor=south west, 
				anchor= north , 
			} ,
			]
			\addplot[color=red,solid,mark=*,line width=2] plot coordinates {
				(1,     0)
				(3.33,  0)
				(6.67,  79.8)
				(10,    81.1)
				(33.3,  81.2)
				(66.6,  81.2)
				(100,   81.2)
                (333.33, 85.4)
                (666.67, 85.4)
                (1000,   85.4)
                (3333.33, 88.5)
                (6666.67, 88.5)
                (10000,   88.5)
			};
			\addplot[color=olive, solid, mark=*, mark size = 2, line width=2] plot coordinates {
				(1,     0)
				(3.33,  0)
				(6.67,  79.8)
				(10,    79.8)
				(33.3,  81.1)
				(66.6,  81.1)
				(100,   81.1)
                (333.33, 81.1)
                (666.67, 81.1)
                (1000,   81.1)
                (3333.33, 81.1)
                (6666.67, 81.1)
                (10000,   81.1)
			};
			\addplot[color=blue, solid, mark=*, line width=2] plot coordinates {
				(1,     0)
				(3.33,  0)
				(6.67,  79.2)
				(10,    79.5)
				(33.3,  79.5)
				(66.6,  79.5)
				(100,   79.5)
                (333.33, 81.2)
                (666.67, 81.2)
                (1000,   85.4)
                (3333.33, 85.4)
                (6666.67, 85.4)
                (10000,   85.4)
			};
		\legend{\Huge{\bf{FuCE}}\\\Huge{\bf{S2E}}\\\Huge{\bf{AFL}}\\}
			\end{axis}
		\end{tikzpicture}
	}
	\vspace{0.3in}
    \caption{Branch coverage obtained on S3C benchmarks after running \emph{FuCE}, \emph{S2E} and \emph{AFL} for two hours}
    \label{fig:coverageOfFuCE}
\end{figure*}

%% file: 6_conclusion.tex
\section{Conclusion}
\label{sec:conclusion}
In our work here, we have identified existing challenges of test-based hardware trojan detection techniques on high-level synthesized design. To this end, we proposed an end-to-end test-generation framework penetrating into deeper program segments at the HLS level. Our results show faster detection of trojans with better coverage scores than earlier methods.  The complete framework for  our proposed \emph{FuCE} test-generation framework has the potential to reinforce automated detection of security vulnerabilities present in HLS designs\cite{farimah2021}. We are investigating the ways in which trojans in RTL/gate level netlist get manifested in HLS using tools like VeriIntel2C, Verilator etc. Future works include exploration of input grammar aware fuzzing and more focus on coverage metrics such as Modified Condition/Decision Coverage (MC/DC)~\cite{codeCoverage} and path coverage.